%% file: main.tex
\documentclass[sigconf, nonacm]{acmart}
\pdfoutput=1
\usepackage{textcomp}
\usepackage{array}
\usepackage{subcaption}
\usepackage{hyperref}
\usepackage{diagbox}
\usepackage{tikz}
\usepackage{changepage}
\usepackage{enumitem}
\usepackage{pifont}
\newcommand{\cmark}{\ding{51}}%
\newcommand{\xmark}{\ding{55}}%
\usetikzlibrary{calc}
\usetikzlibrary{fit}
\usetikzlibrary{positioning}
\usetikzlibrary{shapes.symbols}
\usetikzlibrary{shapes.geometric}
\usepackage{pgfplots}
\newcolumntype{C}[1]{>{\PreserveBackslash\centering}p{#1}}
\newcolumntype{R}[1]{>{\PreserveBackslash\raggedright}p{#1}}
\newcommand{\PreserveBackslash}[1]{\let\temp=\\#1\let\\=\temp}
\usetikzlibrary{shapes, arrows, positioning, fit, calc, decorations.markings,
decorations.pathmorphing, shadows, backgrounds, positioning, patterns}
\tikzset{%
  cascaded/.style = {%
    general shadow = {%
      shadow scale = 1,
      shadow xshift = -1ex,
      shadow yshift = 1ex,
      draw,
      thick,
      fill = white},
    general shadow = {%
      shadow scale = 1,
      shadow xshift = -.5ex,
      shadow yshift = .5ex,
      draw,
      thick,
      fill = white},
    fill = white,
    draw,
    thick,
    minimum width = 0.5cm,
    minimum height = 0.5cm}}

\newcommand\vldbavailabilityurl{}

\begin{document}
\title{How to use Persistent Memory in your Database}

\author{Dimitrios Koutsoukos, Raghav Bhartia, Ana Klimovic, Gustavo Alonso}
\affiliation{%
  \institution{Systems Group, Department of Computer Science}
  \city{ETH Zurich}
  \state{Switzerland}
}
\email{{dkoutsou, rbhartia, aklimovic}@ethz.ch, alonso@inf.ethz.ch}
\begin{abstract}
Persistent or Non Volatile Memory (PMEM or NVM) has recently become
commercially available under several configurations
with different purposes and goals. Despite the attention to the
topic, we are not aware of a comprehensive empirical analysis
of existing relational database engines under different PMEM configurations.
Such a study is important to understand the performance implications of
the various hardware configurations and how different DB engines can benefit
from them. To this end, we analyze
three different engines (PostgreSQL, MySQL, and SQLServer) under common
workloads (TPC-C and TPC-H) with all possible PMEM configurations
supported by Intel's Optane NVM devices (PMEM as persistent memory in AppDirect mode
and PMEM as volatile memory in Memory mode). Our results paint a complex picture and are
not always intuitive
due to the many factors involved. Based on our findings, we provide insights on how the different engines behave with PMEM and which configurations and queries
perform best. Our
results show that using PMEM as persistent storage usually speeds up query
execution, but with some caveats as the I/O path is not fully
optimized. Additionally, using PMEM in Memory mode does not offer any performance
advantage despite the larger volatile memory capacity.
Through the extensive coverage of engines and parameters, we provide an important
starting point for exploiting PMEM in databases and tuning
relational engines to take advantage of this new technology.
\end{abstract}

\maketitle

\begingroup
\renewcommand\thefootnote{}\footnote{\noindent
This work is licensed under the Creative Commons BY-NC-ND 4.0 International License. Visit \url{https://creativecommons.org/licenses/by-nc-nd/4.0/} to view a copy of this license. Copyright is held by the owner/author(s).
}\addtocounter{footnote}{-1}\endgroup

\ifdefempty{\vldbavailabilityurl}{}{
\vspace{.3cm}
\begingroup\small\noindent\raggedright\textbf{PVLDB Artifact Availability:}\\
The source code, data, and/or other artifacts have been made available at \url{\vldbavailabilityurl}.
\endgroup
}

\section{Introduction}
I/O overhead is one of the main bottlenecks in database engines. The growing DRAM capacity over the years has helped but data
sizes keep increasing~\cite{lagar2019software}, making DRAM capacity
insufficient and often too expensive.
To alleviate the memory pressure and to accelerate I/O, Persistent or Non Volatile
Memory (PMEM or NVM) has been proposed as a solution. NVM is both cheaper and
has a higher capacity than DRAM, it is byte-addressable, and it persists data.
The price is higher latency and lower bandwidth.
Database researchers have
extensively studied how to integrate NVM into various layers of a
DBMS~\cite{arulraj2017build,arulraj2019non,arulraj2015let,arulraj2019multi,van2018managing}.
Nevertheless, most of these studies are based on simulating PMEM since they
were done before it was commercially available. The picture has vastly
changed since the release of
Intel\textregistered Optane\textcopyright DC Persistent Memory
Module~\cite{inteloptane}, the first public commercial implementation of
persistent memory. Intel\textregistered Optane\textcopyright DC
Persistent Memory can be configured in Memory, AppDirect or Mixed mode. In
Memory mode it operates as a volatile memory extension, in AppDirect mode it
serves as byte-addressable persistent memory, and finally in Mixed mode
part of it runs in AppDirect mode and the other part in Memory mode.

Optane has sparked numerous studies that benchmark
and explain the behaviour of the
hardware~\cite{izraelevitz2019basic,patil2019performance,peng2019system,weiland2019early}.
These studies demonstrated the capabilities of Intel Optane and
illustrated its basic characteristics compared to DRAM and SSDs. For example,
they showed that for sequential workloads PMEM has
comparable latency to DRAM while it is considerably slower for
random accesses. They also demonstrated the asymmetric behaviour of PMEM
between read and write bandwidth and that interference between different workloads
drastically affects performance.

However, DBMSs are complex systems that behave very differently with
the workload or the amount of data. Additionally, databases have many
knobs that can be altered, which can affect performance significantly.
Although the database community has started to evaluated PMEM on relational
workloads~\cite{daase2021maximizing,bother2021drop,wu2020lessons}, these studies
give only part of the picture as they provide only high-level conclusions and
are mostly based on micro- or small-scale benchmarks. Furthermore, they do not
experiment with different database knobs and they do not correlate database
characteristics with PMEM behaviour. To this end, in this paper we provide what
to our knowledge is the first extensive analysis relational database engines using
Intel\textregistered Optane\textcopyright DC Persistent
Memory: we run OLAP and OLTP
workloads (TPC-H and TPC-C) on three DBMSs
(PostgreSQL, MySQL, and SQLServer) under various PMEM configurations allowed by
the hardware (i.e., PMEM as volatile and persistent
memory). We modify the database configurations to put I/O as much as possible
in the critical path to magnify the hardware differences. We also experiment with
 database knobs to see their effect on performance. Additionally,
we give details on how query plans and different operators behave on PMEM.

Our
results provide a complex picture that shows that using PMEM does not always
translate into better performance, despite its hardware
advantage. Although PMEM is faster than SSDs, the I/O
path is not fully optimized since there is no OS prefetching and the CPU is involved
in I/O when using PMEM as persistent memory. Furthermore, in systems where
there is a lot
of resource contention and mixed workloads, PMEM experiences a large performance
drop. That makes SSDs competitive for several scenarios, e.g. sequential
queries with high selectivity. Similarly, although PMEM offers extra volatile
memory capacity in Memory mode, this does not translate to a performance
benefit. As of today, and based on these results, it seems that the best use of
PMEM is as faster (but more expensive) storage. However, there are many caveats and researchers
should carefully tune their engines and take into account the nature of the
workloads to leverage the new hardware.

\section{Background}
%
\subsection{Persistent Memory}
Persistent memory, also known as non-volatile or storage class memory,
is an emerging class of memory technology combining the byte-addressability
and low latency of DRAM with the persistence and high capacity of disks. The
technology is now available in the Intel\textregistered
Optane\textcopyright DC Persistent Memory Module~\cite{inteloptane}
(referred to as PMEM in the rest of the paper). It is based on 3D
Xpoint technology, which is faster and more expensive than NAND Flash but slower
and cheaper than DRAM. PMEM comes in a DIMM form factor in 128, 256, or 512 GB
sizes and it is attached to a memory channel.

\begin{figure}[t]
  \centering
  \input{figures/architecture}
  \caption{Socket topology}\label{fig:socket-topology}
    \vspace{-1em}
\end{figure}
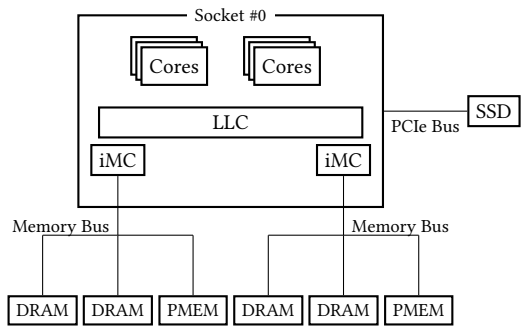
\begin{figure}[t]
    \begin{subfigure}[t]{0.16\textwidth}
        \centering
        \input{figures/memorymode}
        \caption{Memory}\label{fig:memorymode}
    \end{subfigure}
    \begin{subfigure}[t]{0.16\textwidth}
        \centering
        \input{figures/storagemode}
        \caption{AppDirect}\label{fig:storagemode}
    \end{subfigure}
    \begin{subfigure}[t]{0.13\textwidth}
        \centering
        \input{figures/mixedmode}
        \vspace{-0.38cm}
        \caption{Mixed}\label{fig:mixedmode}
    \end{subfigure}
    \caption{Memory hierarchy for different PMEM modes}\label{fig:memhierarchy}
    \vspace{-2em}
\end{figure}
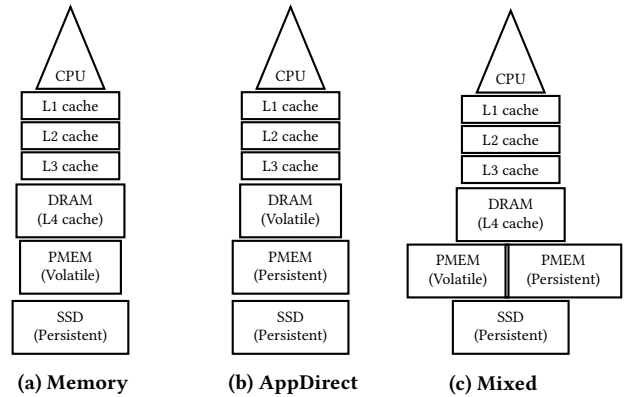

Every memory channel is connected directly to the CPU
through an integrated Memory Controller (iMC) (Figure~\ref{fig:socket-topology}). The iMC maintains
read and write pending queues for the PMEM DIMMs and it is situated within
Intel's asynchronous DRAM refresh domain, which ensures persistence on power
failure. Every iMC has at most 3 memory channels and every CPU socket has two
iMCs, resulting in a maximum memory capacity of 6TB for a 2-socket server with
12 memory channels. The access granularity from the iMC to PMEM is 64 bytes,
however PMEM's
internal access granularity is 256 bytes. That may lead to read or write
amplification in different access patterns. For example, in sequential access
patterns, PMEM has the highest performance and it is only 2x slower than DRAM,
since one internal access serves four external accesses.
PMEM can be used in 3 modes: \textit{Memory Mode}, \textit{AppDirect
Mode}, and \textit{Mixed Mode} (Figure~\ref{fig:memhierarchy}).
We denote in parenthesis how the OS treats DRAM
or PMEM in different modes.

In \textit{Memory Mode}, PMEM is used as volatile memory and
DRAM becomes an L4 direct-mapped cache~\cite{intelfactsheet} transparent
to the application. The iMC manages data traffic between DRAM and PMEM. If a read
request results in a DRAM cache hit, the memory controller serves the read. In
case of a DRAM cache miss, it issues a second read to PMEM. Therefore, for a
cache hit the application experiences only the DRAM latency, whereas for a cache
miss, it experiences the latency of both PMEM and DRAM. Applications can use
PMEM in this mode with no changes in their source code.

In \textit{AppDirect Mode}, PMEM acts as a persistent storage device much as an SSD
or HDD disk. It can be configured as interleaved or standalone
regions. When used as an interleaved region, PMEM provides striped read/write
operations that offer increased throughput. The supported interleaved size is 4
KB. In this mode, PMEM contains one or more namespaces, which are similar to
hard disk partitions. A namespace can be configured in \texttt{fsdax},
\texttt{devdax}, \texttt{sector}, or \texttt{raw} mode. \texttt{Fsdax} provides a
file system with direct access. If we mount on the device a DAX-aware file
system, we can execute load and stores from/to PMEM without using the page cache
or any other OS intervention. This is the mode that is recommended by Intel.
\texttt{Devdax} presents the device as a character
device, exposed to the OS as a single file. Therefore, no filesystem can be
mounted on it and read/write system calls are not supported. \texttt{Sector}
mode configures the storage as a block device on which any non-DAX file system
can be mounted. Finally, \texttt{raw} mode presents the device as a memory disk
without any support for mounting a DAX filesystem.

\textit{Mixed Mode} is a combination of Memory and App Direct modes. The entire
DRAM is again configured as an L4 cache. The user can configure the percentage
of PMEM that the system will use as volatile memory or disk, however the minimum
memory DRAM to PMEM ratio recommended by Intel is 1:4.

\subsection{Related work}
Even before the first public hardware implementation of Non-Volatile Memory
(NVM),
researchers and companies were adapting DBMS components or tailoring DBMSs to
NVM characteristics. Chatzistergiou et al.~\cite{chatzistergiou2015rewind}
developed a library that manages transactional updates in data structures
designed for NVM and Wang et al.~\cite{wang2014scalable} came up with a
distributed logging protocol that leverages NVM.
Similarly, Arulraj and Pavlo adapted different parts
of a database engine to NVM, such as the buffer
manager~\cite{zhou2021spitfire}, and the storage and
recovery~\cite{arulraj2016write, arulraj2015let} protocol. With their findings,
they developed a DBMS, Peloton~\cite{pavlo2017self, peloton}, built to leverage NVM principles and characteristics.
Similarly, SAP HANA adapted their engine to integrate NVM~\cite{andrei2017sap} and
Liu et al.~\cite{liu2021zen} built a log-free OLTP engine for NVM. Finally, van
Renen et al.~\cite{van2018managing} evaluate how DBMSs perform when they use either NVM
exclusively or a hybrid approach with a page-based DRAM cache in front of NVM.
In addition to these efforts building new engines or parts of it to take
advantage of PMEM, there has been considerable work on algorithms such as
index joins~\cite{psaropoulos2019bridging}, or data structures such as
B+-Trees~\cite{chen2015persistent}, hash tables~\cite{lu2020dash}, and range
indexes~\cite{lersch2019evaluating}.

In most of the above efforts, the authors were simulating NVM using software
tools~\cite{dulloor2014system} to benchmark the resulting implementations.
However, with the introduction of Intel Optane Persistent Memory, it is possible
to execute the workloads on actual hardware. This has led to several performance
analyses appearing in the last years.
Outside of the database context,
Swanson et. al~\cite{izraelevitz2019basic}
provided the first comprehensive performance measurements of
Intel Optane Persistent Memory including read/write latency with sequential and
random access patterns as well as the effect of factors such as the number of threads on different PMEM
modes (Memory-mode and AppDirect-mode). Based on their findings, they propose best
practices on how to leverage PMEM. This
study was followed by others~\cite{peng2019system,yang2020empirical} providing
more in-depth analysis and elaborating on the insights of Swanson et.
al. In HPC, a number of
studies~\cite{patil2019performance,weiland2019early} have shown promising results
in hybrid DRAM-NVM configurations. Using such a configuration, the applications
experience only a minimal slowdown and they can scale up to a larger amount of
data, due to the larger volatile memory capacity.

In databases, the focus until now has been mainly on OLAP workloads.
Benson et. al.~\cite{daase2021maximizing,bother2021drop} use microbenchmarks
or more sophisticated benchmarks (Star Schema, TPC-H) to provide best practices
for using Intel Optane Persistent Memory and to investigate if it is a valid
replacement for NVMe SSDs. In their experiments, they only use the AppDirect mode (both
with and without \texttt{fsdax}). Shanbhag et. al~\cite{shanbhag2020large}
also evaluate only the AppDirect mode in an OLAP context.
Complementary to that, Wu et. al~\cite{wu2020lessons} have run microbenchmarks,
TPC-H, and TPC-C on SQL server 2019 using PMEM in Memory and
AppDirect mode (with and without \texttt{fsdax}) and they
present high-level conclusions on how DBMS can leverage PMEM. Additionally,
Renen et al.~\cite{van2019persistent} measure the bandwidth and latency of
Intel Optane Persistent Memory and based on their findings they tune log
writing and block flushing.
Compared to all the aforementioned efforts, besides focusing only on more
sophisticated benchmarks (TPC-H and TPC-C) instead of microbenchmarks, we
also evaluate three different DBMSs in both AppDirect and Memory mode by
altering various configurations and database knobs. In TPC-H, we are the
first to show how query plans, specific operations, and system statistics are
affected by the usage of PMEM. In TPC-C, we do a deep exploration of the
available database knobs and how they affect performance.
\section{Experimental setup}
\begin{table}[t]
    \begin{tabular}{c c}
        \toprule
        \textbf{Component} & \textbf{Specs} \\
        \midrule
        Sockets & 2 \\
        CPU & Intel(R) Xeon(R) Gold 6248R \\
        Microarchitecture & Cascade Lake \\
        Cores & 48 physical (96 logical) \\
        L1 cache & 64KB \\
        L2 cache & 1MB \\
        L3 cache & 36MB \\
        RAM & 128GB (16GB DDR4 @ 2666 MHz $\times$ 8) \\
        PMEM & 512GB (128GB $\times$ 4) \\
        SSD & KIOXIA KPM6XRUG1T92 (2TB) \\
        \bottomrule
    \end{tabular}
    \caption{Server specifications}
    \label{tab:specs}
    \vspace{-3em}
\end{table}

\subsection{System specification}

All the experiments in this paper were conducted on a dual-socket server with
the specs mentioned in Table~\ref{tab:specs} and the socket topology shown in
Figure~\ref{fig:socket-topology}. Each CPU socket has two memory controllers and three
channels per controller. The DRAM and Intel Optane DIMMs are installed in a
1-1-1 topology. We use Debian 10 (kernel 5.10.46). We disable the turbo mode and
we set the CPU power governor to performance mode with 2GHz as the maximum
frequency, so that we can have reproducible experiments with minimum variation.
We refer to \textit{Default Mode} as the configuration that does not use PMEM at all
and, unless otherwise mentioned, to AppDirect mode as the configuration that mounts
PMEM in AppDirect mode with \texttt{fsdax} enabled. We collect system
statistics with the Intel Vtune Platform Profiler~\cite{intelvpp} that provides metrics
related to CPU, memory, storage, and I/O. To get more accurate statistics, we
modify the source code of the profiler to increase the sampling hardware counter
frequency.

\begin{table}[b]
    \begin{tabular}{l c}
        \toprule
        \textbf{Memory type} & \textbf{Latency [us]} \\
        \midrule
        DRAM & 0.07 \\
        PMEM-volatile & 0.09 \\
        PMEM-storage (random) & 9.31 \\
        PMEM-storage (sequential) & 9.23\\
        SSD (random) & 68.6 \\
        SSD (sequential) & 65.8 \\
        \bottomrule
    \end{tabular}
    \caption{Latency measurements for different memory types in
    our server}
    \label{tab:latency}
    \vspace{-3em}
\end{table}

\begin{table}[t]
    \begin{tabular}{R{4.8cm} C{1.4cm} C{1.4cm}}
        \toprule
        \textbf{Memory type} & \textbf{Read bw [MB/s]} & \textbf{Write bw [MB/s]} \\
        \midrule
        DRAM & 76720 & 57388 \\
        PMEM-volatile & 75303 & 32094 \\
        PMEM-storage (random, 8 threads) & 6128 & 1807 \\
        PMEM-storage (sequential, 8 threads) & 8551 & 3694\\
        PMEM-storage (random, 1 thread) & 1829 & 534 \\
        PMEM-storage (sequential, 1 thread) & 2449 & 615 \\
        SSD (random, 8 threads) & 470 & 293 \\
        SSD (sequential, 8 threads) & 504 & 477 \\
        SSD (random, 1 thread) & 160 & 157 \\
        SSD (sequential, 1 thread) & 191 & 166\\
        \bottomrule
    \end{tabular}
    \caption{Bandwidth measurements for different memory types in
    our server}
    \label{tab:bandwidth}
    \vspace{-3mm}
\end{table}

\subsection{Basic microbenchmarks}

Before evaluating database engines, we measured
the latency and the bandwidth of our system
for the different memory types (Tables~\ref{tab:latency} and
\ref{tab:bandwidth}). For volatile memory (DRAM and
PMEM in Memory
mode) we use the Intel Memory Checker Tool~\cite{mlc}, which gives us the
intrasocket idle memory latencies. For measuring latency in
storage (SSD and PMEM in AppDirect mode) we use ioping~\cite{ioping}. Finally, for
measuring bandwidth in storage we use fio~\cite{fio}. For storage bandwidth measurements
we follow a similar methodology to Swanson et. al~\cite{izraelevitz2019basic}.
We perform two sets of measurements, one with 1 thread and one with 8 threads,
to observe how bandwidth scales with the number of threads.
We generate 512MB files and every thread reads 1 file in 4KB blocks. We bypass
the page cache and we perform synchronous I/O. For writes, we perform a
\texttt{fsync()} call after every 4KB write. As we can observe, PMEM as volatile
memory has 23\% higher latency. The read bandwidth is 2\% less and the write
bandwidth is more than 2$\times$ lower than DRAM. For 1 thread PMEM AppDirect has more
than 10$\times$ more read bandwidth for both random and sequential accesses and
around 5$\times$ more write bandwidth than SSD. For 8 threads PMEM has
13-16$\times$ higher read bandwidth and 3-6$\times$ higher write bandwidth for
random and sequential accesses, respectively, than SSD. Finally, we notice a
large gap in bandwidth between using PMEM in Memory or AppDirect mode, but this
is due to the overhead of the OS and the filesystem.

\subsection{Configurations tested}
\begin{table}[b]
    \begin{tabular}{|c|*{5}{c|}}\hline
    \backslashbox{Benchmark}{Mode}
    &\makebox[2.5em]{Default}&\makebox[2.5em]{Memory}&\makebox[3.2em]{AppDirect}
    &\makebox[2.5em]{Mixed}\\\hline
    TPC-H & \cmark & \cmark & \cmark & \xmark \\
    TPC-C & \cmark & \cmark & \cmark & \xmark \\
    \hline
    \end{tabular}
    \caption{Benchmarks and modes run for all the different DBMSs for
    PostgreSQL, MySQL and SQLServer}
    \label{tab:overview}
    \vspace{-2em}
\end{table}
We now present the different configurations that we run for the TPC-H and TPC-C
benchmarks. For both TPC-H and TPC-C, we pin the database processes to socket 0
to avoid remote NUMA access to better interpret the results. Before executing
each query or workload we clear the page and the database cache. For TPC-C, we
use 1000 warehouses for all experiments, unless otherwise stated. We configure
each user to use all the warehouses because this ensures
an even distribution of virtual users across warehouses and avoids spending a
lot of time in lock contention. We set the warmup and running time to 3 minutes.
That is sufficient to load the buffer pool and stabilize the database activity.
Additionally, we have observed that we manage to reach the maximum tpmC
by using only one of the two PMEM DIMMs available in the socket. Therefore for
TPC-C unless otherwise mentioned, the AppDirect mode uses one PMEM DIMM.
Although we run the Memory mode for TPC-C for many configurations, in
most cases its performance is almost identical to the Default case as requests
are served by the L4 DRAM cache and PMEM is not in the critical path. We
therefore do not report any results involving the Memory mode. Finally, we do not
run any experiments for Mixed mode, since we do not meet the minimum volatile
memory ratio of 1:4 in our setup, which Intel suggests for PMEM to be
cost and performance-effective over a DRAM only solution. We have 128GB of RAM
and 512GB of PMEM. That leaves no PMEM capacity to be configured in AppDirect
mode. Table~\ref{tab:overview} summarizes the different modes and benchmarks that have been
run for all DBMSs considered.

\section{PostgreSQL}\label{sec:postgres}
\subsection{TPC-H}
\begin{figure}[b]
  \begin{adjustwidth*}{}{-1.5em}
  \centering
  \input{plots/postgres-tpch-runtime}
  \end{adjustwidth*}
  \caption{Running time (log-scale) for TPC-H SF-100 on PostgreSQL for different system
  configurations}\label{fig:tpch-time-postgres}
\vspace{-1em}
\end{figure}
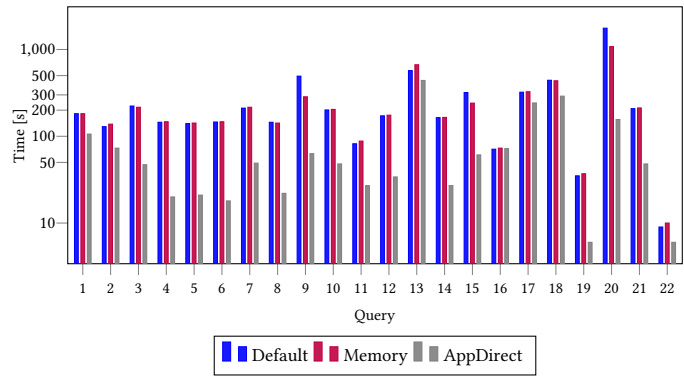
\subsubsection{Setup}\label{subsec:olap-postgres}
We set
the buffer cache to 16GB and the effective cache size to 48GB. That way
all the caches fit in socket 0 for all modes and therefore I/O is in the
critical path as much as possible.
We use PostgreSQL version 13.2. After importing the data, we
add indexes and foreign keys with triggers disabled. Postgres does not have a
prefetcher on its own and relies on the OS for this purpose.

\subsubsection{General observations\label{sec:postgres-general-tpch}}
The running times for the 22 TPC-H queries for PostgreSQL for the
Default, Memory, and AppDirect modes are shown in Figure~\ref{fig:tpch-time-postgres}.
The AppDirect mode is consistently faster than the Default mode. This happens
because PMEM has lower latency than SSDs and, thus, all I/O is faster. This is more obvious in queries that
have large sequential scans (queries  1, 3, 6, 7, 12, 16, 22). Although
the lower latency does provide a large advantage, it is not as
large as we would expect (around 8-10$\times$ according to Table \ref{tab:latency}).
That happens because the AppDirect mode does not use the OS page cache and
prefetcher. Therefore, in queries where these components are effectively hiding the I/O latency, the
advantage is not large (e.g., query 1), contrary to queries where they do
not play a big role (e.g., query 6). Another reason for not seeing all potential advantages is that, in AppDirect mode, the CPU is involved in I/O, whereas in
Default mode a parallel worker thread is scheduled to deal with I/O.
To verify how much the number of I/O threads affects the overall performance, we
also run the AppDirect and Default modes using only one core on socket 0 and we
observe that the performance difference between them reduces drastically.
However, the AppDirect mode still performs better because, even with one thread,
PMEM has a 10$\times$ higher read bandwidth (Table \ref{tab:bandwidth}). Nevertheless, there are situations when AppDirect mode works much better: for parallel
index lookups the latency difference is
about 10$\times$ faster for the AppDirect mode (queries 2, 4, 5). Especially when
the tables are large, the lookup access size is in the order of 8KB and, although
the accesses are random, they have the same latency as a sequential scan for
PMEM (queries 8, 9, 10, 19, 20, 21).

\begin{figure}[t]
  \begin{adjustwidth*}{}{-1.5em}
  \centering
  \input{plots/postgres-dram-read}
  \end{adjustwidth*}
  \caption{DRAM average read bandwidth on PostgreSQL for different system
  configurations}\label{fig:tpch-dram-read-postgres}
  \vspace{-2mm}
\end{figure}
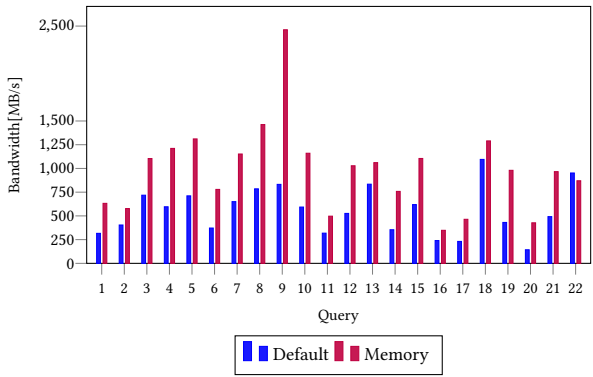
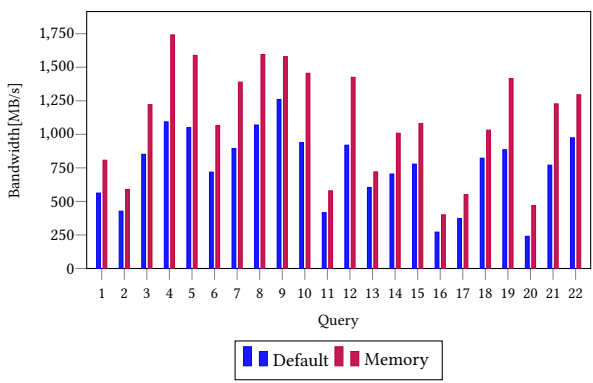
\begin{figure}[b]
  \begin{adjustwidth*}{}{-1.5em}
  \centering
  \input{plots/postgres-dram-write}
  \end{adjustwidth*}
  \caption{DRAM average write bandwidth on PostgreSQL for different system
  configurations}
  \vspace{-2em}
  \label{fig:tpch-dram-write-postgres}
\end{figure}
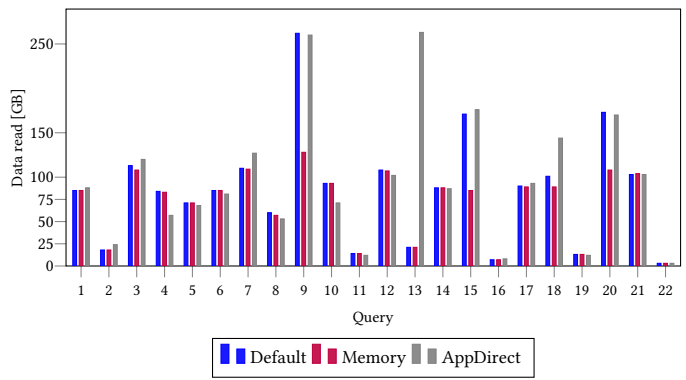
\begin{figure}[t]
  \begin{adjustwidth*}{-1.2em}{}
  \centering
  \input{plots/postgres-read-storage}
  \end{adjustwidth*}
  \caption{Data from storage read for TPC-H SF-100 on PostgreSQL for different system
  configurations}\label{fig:tpch-read-storage-postgres}
  \vspace{-2em}
\end{figure}

The Memory mode has slightly worse running times
for most queries. To understand where the overhead comes from, in
Figures~\ref{fig:tpch-dram-read-postgres} and
\ref{fig:tpch-dram-write-postgres} we show the DRAM read and write
bandwidth consumed by each query for
the Default and Memory modes, respectively. For all the queries, we read and
write more data from DRAM in Memory Mode compared to the Default mode. For
reads, that happens because a DRAM (L4 cache) read miss in Memory mode results
in a read from PMEM, which consequently leads to a DRAM write. Another reason
for the increased DRAM reads is that during a DRAM write, the system has to read
the dirty lines and flush them to PMEM. Due to
prefetching by the iMC, the DRAM writes are more than the L4 cache misses.
Finally, another reason for the increased DRAM bandwidth for both reads and
writes are the collisions in the L4 direct-mapped cache, since multiple PMEM
memory lines can map to the same DRAM memory line.
We verify the last assumption by allocating
64GB instead of 256GB of volatile memory in Memory mode. This way,
two PMEM memory lines always map to different
DRAM memory lines. This decreases the cache conflicts and the DRAM read/write
bandwidth and we observe that the Memory mode has the same running time as the
Default mode. We also see in Figure~\ref{fig:tpch-read-storage-postgres} that the
Memory mode consistently reads
fewer or equal amounts of data because of the larger page cache available,
whereas the AppDirect mode consistently reads an equal or bigger
amount of data because of the disabled page cache. Exceptions to that are
queries 4, 8, 10 and 11, because of the inability to take advantage of the OS
cache and prefetching.

We now analyze the differences between the Default and the Memory mode in more
detail. Queries 1, 4, 5, 6, 12, 14 and 17 are less than 2\% slower. This
happens because their working set is less than 64GB and therefore most of the
accesses are served by the L4 DRAM cache.
On the other hand, queries 2, 11, 13, and 19 are slower
by more than 5\%, because their working set is larger than the L4 and/or the
OS page cache. Finally, queries 9, 15, and 20 are faster in
the Memory mode. These queries benefit from the larger OS page cache that the
OS can allocate, since socket 0 with PMEM in Memory mode has 256GB of volatile
memory, compared to the Default mode which has 64GB.

\subsubsection{Query-by-query}
In this section, we take a more detailed look at the queries to gain
additional insights. Note that we only compare the Default with the AppDirect
mode.

Query 1 has a
parallel sequential scan and a parallel aggregation of \texttt{lineitem}. The
selection predicate is satisfied by 98\% of the rows. In AppDirect mode, 85 out of
the 106 seconds are spent in computing the aggregations, and only 17 seconds are
spent in scanning the data. PMEM has a latency of 10us for 4KB application
reads~\cite{intellatency}. Therefore, for the block size of Postgres (8KB),
the latency should be less than 20us. We can confirm that because we have 10
parallel workers with an overall bandwidth of 4-5 GB/s which explains the 17
seconds needed to scan the 86GB table. In contrast, in
Default mode, SSDs have a 10x latency, but only 5x of the time is spent in scanning
because of the OS page cache and prefetching.


In query 3 the Default mode spends the majority of the time scanning the
\texttt{lineitem} table. The time is larger than query 1 because the prefetcher
does not work so effectively. That happens because the inter-arrival time of
read calls is lower as fewer operations are performed in higher-level operators.
For the AppDirect mode, the scan time increases to 19 seconds compared to query 1.
This is because parallel workers are also involved in writing data to PMEM
during the join and sorting of the query. It is well known that writing in
Intel Optane requires more CPU and memory resources and provides lower
bandwidth~\cite{daase2021maximizing, izraelevitz2019basic}

In query 5, there is a multibatch parallel hash join where the table
\texttt{orders} is scanned. The join does not start executing until the entire
table has been scanned. Therefore, the inter-arrival times of read requests are
very close and the prefetcher does not work effectively in the Default mode.


Query 13 has an index-only sequential scan on \texttt{customer}, which takes the
same time in both modes due to prefetching in the Default mode. There is also an
index scan in \texttt{orders}, which takes approximately the same time in the
AppDirect and Default mode. That happens because the index is not correlated, and
therefore the database can scan more than one block at the same time. This
leads to a large working set that is larger than the buffer cache, which
subsequently leads to many reads from PMEM in AppDirect mode. The amount of data
read in AppDirect mode is 12$\times$ more than that in the Default mode. However,
AppDirect mode still performs better, because PMEM bandwidth is 6$\times$ larger
and it does not involve the synchronous read from DRAM as the Default mode.

Query 14 involves a hash join between \texttt{lineitem} and \texttt{part} with
lineitem on the build side. The hash table has to be built completely before the
rest of the query plan is executed and therefore we have a sequential scan on
\texttt{lineitem}. For the scanning, 6 workers are allocated. The SSD reaches
its maximum read bandwidth, but this is not the case for PMEM, which does not
reach its maximum read bandwidth even with 9 workers.
In query 15, we observe
the asymmetric behaviour of PMEM as we have to read \texttt{lineitem}
twice. The maximum read bandwidth is 4 GB/s and the maximum write bandwidth is 1
GB/s. To maintain the price/performance ratio, we need to increase the working
memory or use a separate storage drive for temporary writes.

Lastly, in query 17 there is a hash join between \texttt{lineitem} and \texttt{part}
with \texttt{lineitem} on the probe side without any filtering. For the
sequential scan of lineitem, prefetching is effective and the main performance
difference is caused by index lookups on lineitem. In such CPU intensive
queries, when CPU operations overshadow I/O requests, prefetching is very
effective in sequential accesses. On the other hand, for random accesses
asynchronous I/O is more beneficial. Therefore, if Postgres adopts asynchronous
I/O, the performance difference between the two modes would be negligible.

\subsubsection{Summary}
The AppDirect mode is largely beneficial for sequential scans that select very
little data. When the majority of the data has to be processed, then the lack of
the page cache and prefetching in AppDirect mode eliminates most of the advantages of PMEM.
The Default mode is also competitive in queries involving CPU-intensive
operations (hash aggregations and joins) when I/O requests
are not the bottleneck. Finally, because the CPU is directly involved in I/O in
AppDirect mode, if we do not use a high number of threads, the bandwidth
utilization is not optimal and the performance is comparable to the Default mode.
The Memory mode is slightly slower than the Default mode, except for queries
that have a large working set and can benefit from the larger OS page cache offered by
PMEM. In addition, the
direct-mapped L4 cache leads to larger DRAM utilization due to conflict misses.
Applications could allocate volatile memory space
contiguously to avoid them.

\subsection{TPC-C}
\subsubsection{Setup}\label{subsubsec:tpcc-postgres}
We use PGTune~\cite{pgtune} to tune PostgreSQL to our hardware configuration. We
set the \texttt{checkpoint\_timeout} to 30 seconds and the
\texttt{max\_wal\_size} to 5GB to force checkpoints very often. It is known that
writes have limited bandwidth in PMEM (as seen in Table~\ref{tab:bandwidth}). By
checkpointing aggressively, we put I/O in the critical path and we can
saturate the PMEM device more easily. This configuration, though not very common,
is useful for meeting a very low downtime requirement. We also use a
large buffer pool to reduce the proportion of reads to writes.
After the warmup time, we
checkpoint and run a \texttt{VACUUM} command to remove potential sources of performance variations.

\begin{figure}[t]
  \begin{adjustwidth*}{}{-1.5em}
  \centering
  \input{plots/postgres-tpcc-users}
  \end{adjustwidth*}
  \caption{tpmC for TPC-C with 1000 warehouses on Postgres for different system
  configurations}\label{fig:tpcc-tpm-postgres}
  \vspace{-1em}
\end{figure}
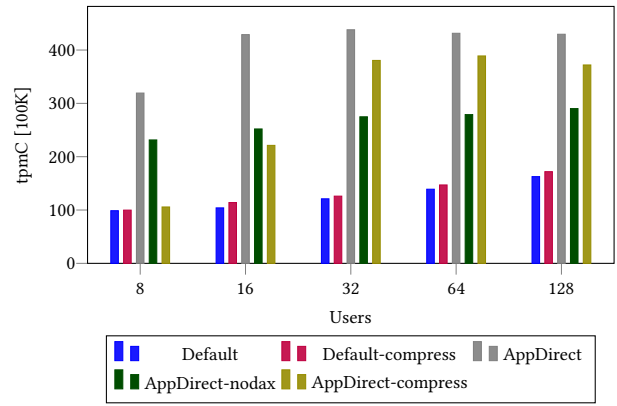

\subsubsection{Different modes and WAL compression}
In Figure~\ref{fig:tpcc-tpm-postgres} we present the result of running TPC-C for
different numbers of concurrent users for different system configurations.
Besides the Default and the AppDirect mode, we run both these modes with an option
that allows as to compress the Write-Ahead Log (WAL) before writing. We denote
them with \texttt{Default-compress} and \texttt{AppDirect-compress} in the figure.
We also run TPC-C in AppDirect mode but with DAX disabled (denoted as
\texttt{AppDirect-nodax} in the figure). As we see, the AppDirect mode reaches a
plateau after 32 concurrent users, while the Default mode slowly increases.
Postgres utilizes the PMEM's bandwidth very efficiently, because the checkpoint,
background, and WAL writer processes use only one thread. Additionally, when the
client workers get involved in flushing dirty pages, we see a slight tpmC drop.

WAL compression is widely used for Postgres to close the gap between CPU and
storage. However, in our case even though it is slightly useful for the Default
mode, it causes a drop in tpmC for the AppDirect mode. The number of transactions
is almost 50\% less for 16 clients. Due to the lower latency of PMEM,
the compression step involving an additional read and write from/to memory
becomes unnecessary. Lastly, because Postgres depends on the OS page cache for
read-ahead and buffering of writes, we enable the page cache in AppDirect Mode with
the no-DAX option. However, as we have mentioned, in AppDirect mode the CPU is
involved in reading/writing and this causes an increase in CPU usage and a drop
in tpmC as we can see in Figure~\ref{fig:tpcc-tpm-postgres}. Therefore, although
Postgres is very effective at hiding the storage latency of HDDs and SSDs, this
is not the case with PMEM and the techniques it uses for HDDs and SSDs hurt performance in PMEM.

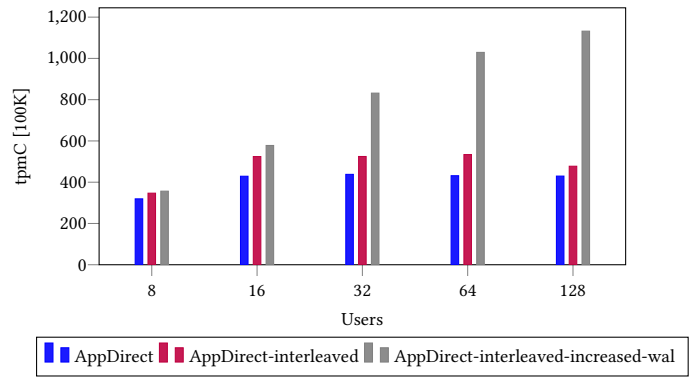
\begin{figure}[t]
  \begin{adjustwidth*}{}{-1.5em}
  \centering
  \input{plots/postgres-tpcc-dimms}
  \end{adjustwidth*}
  \caption{tpmC for TPC-C on Postgres using different
  number of DIMMs}\label{fig:tpcc-dimms-postgres}
\end{figure}

\subsubsection{Interleaved vs non-interleaved}
We compare the AppDirect mode using only one PMEM DIMM with the AppDirect Mode using
both PMEM DIMMs (denoted with AppDirect-interleaved) in
Figure~\ref{fig:tpcc-dimms-postgres}. The interleaving access block size is 4KB
and any read/write for Postgres happens in multiples of 8KB. Therefore, when
using PMEM in interleaved mode, we would expect a large increase in tpmC, since
the access latency is halved and the max bandwidth is doubled. However, as we
see in the figure, the increase in tpmC is minimal. For a small number of clients,
that happens because the workload is already CPU-bound even when using a single
PMEM DIMM. For a larger number of clients, the bottleneck is the checkpointing
process. Thus, we increase the \texttt{max\_wal\_size} to reduce the number of
checkpoints and the tpmC for 128 clients is more than 2x higher and is bounded
by the max bandwidth of the interleaved setup.

\begin{figure}[b]
  \begin{adjustwidth*}{-1.5em}{}
  \centering
  \input{plots/postgres-tpcc-stores}
  \end{adjustwidth*}
  \caption{tpmC for TPC-C on Postgres using PMEM as WAL or data
  store}\label{fig:tpcc-stores-postgres}
  \vspace{-1em}
\end{figure}
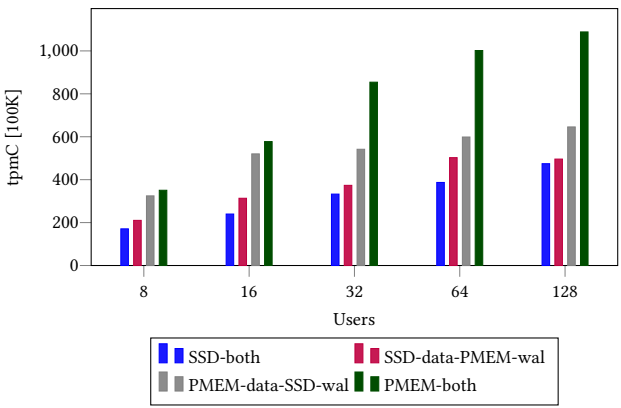

\subsubsection{Data and WAL placement}
Finally, we evaluate the effectiveness of PMEM as data or a WAL store. We
present the results in Figure~\ref{fig:tpcc-stores-postgres}. We put both the
WAL and the data in SSD (\texttt{SSD-both}), the data in SSD and the WAL in PMEM
(\texttt{SSD-data-PMEM-wal}), the data in PMEM and the WAL in SSD
(\texttt{PMEM-data-SSD-wal}), and both the data, and the WAL in PMEM
(\texttt{PMEM-both}). For a small number of clients, when we place the WAL in PMEM
instead of the SSD, we see only a slight increase in tpmC. That means that
buffered I/O does not significantly affect performance in log writing. However,
the gap increases when we increase the clients due to the bandwidth limitations
of the SSD. On the other hand, when we place the data in PMEM instead of the
SSD, we see a large increase in tpmC, even for a small number of clients. We
therefore conclude that PMEM should be used as a data rather than just as a WAL store,
because databases can utilize PMEM's read bandwidth more effectively than its
write bandwidth.

\subsubsection{Summary} We see that compressing the WAL as well as placing it in
faster storage (e.g. PMEM) does not offer us a large performance advantage.
In general, although TPC-C involves many I/O writes and reads, these are
still done on the same data group and that makes the workload more CPU than I/O
bound. Finally, Postgres utilizes the PMEM bandwidth very effectively in DAX
mode but, when the page cache is enabled, the tpmC drops drastically. This
happens because the PMEM latency is low compared to that of SSDs/HDDs and
since Postgres relies on the OS for prefetching, keeping/updating the page cache
is an additional overhead.

\section{MySQL}
\subsection{TPC-H}
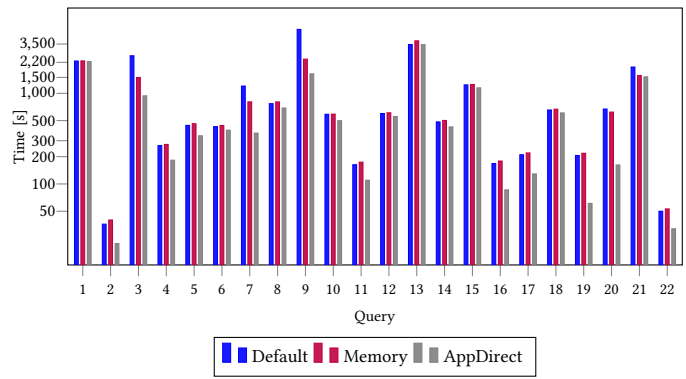
\begin{figure}[b]
  \vspace{-1em}
  \begin{adjustwidth*}{-1.5em}{}
  \centering
  \input{plots/mysql-tpch-runtime}
  \end{adjustwidth*}
  \caption{Running time for TPC-H SF-100 on MySQL for different system
  configurations}\label{fig:tpch-time-mysql}
\end{figure}

\subsubsection{Setup}
We use MySQL version 8.0.23. We set the
buffer pool again to 16GB for the reasons stated in
Section~\ref{subsec:olap-postgres} and to have the same parameters across
databases. MySQL does not support multiple threads per query, except for
particular type of queries, e.g., \texttt{SELECT COUNT(*)}~\cite{mysqlnonparallel}.

\subsubsection{General observations}
In Figure~\ref{fig:tpch-time-mysql}, we present the runtimes for TPC-H SF 100
for MySQL for different configurations.
Regarding the query times, we observe similar trends to Postgres. The difference
between the AppDirect and the Default mode is not as big, because we do not reach
the bandwidth limitations of the SSD with a single thread. In addition, compared
to Postgres, MySQL has two main differences. First, it prefers nested loop joins
over hash or merge joins. Therefore, in queries involving joins, there are
random accesses with large block sizes (i.e. 16KB), where PMEM has similar bandwidth
to sequential accesses. In contrast, prefetching
and the OS page cache cannot help in Default mode for random accesses. Second, MySQL
stores primary clustered indexes together with the data and secondary indexes
separately. That puts PMEM at a disadvantage, because an access to a secondary
index needs to locate the data in the tables and in the Default mode part of
these accesses are served by the OS page cache.

In Memory mode queries 3, 7, 9, 20, and 21 have a smaller running time than the Default
mode, because of the larger page cache available. All these queries have
a large working set and the data does not fit in the
buffer cache. For the other queries, the two modes are either comparable or the
Memory mode is slightly worse for the reasons already mentioned in
Section~\ref{sec:postgres-general-tpch}.

\subsubsection{Query-by-query}
We now take a deeper look into a subset of the queries . We again compare only
the AppDirect and the Default modes, as those are the
ones that have the largest runtime difference. In query 1, most of the time is
spent on aggregation and filtering. The scan time of \texttt{lineitem} in the
Default mode is 436s and in AppDirect mode it is 392s. The small performance difference
has two reasons. First, prefetching and page caching by the OS is
very effective for sequential scans of large tables. Second, the average block
size is around 100KB for both modes. The increase in response time for larger
block sizes is sublinear in SSDs due to the inherent parallelism. On the other
hand, in AppDirect mode the CPU is involved in reading, and the response time is
exactly linear.

In query 3, there is a table scan on \texttt{customer} and index lookups on
\texttt{orders} that use a secondary index. The cost of a lookup is 140us for
the AppDirect mode and 250us for the Default mode. The difference is not as large
as we would expect based on the hardware, but the Default mode has the advantage
of prefetching and the OS page cache. Additionally, PMEM has a larger latency when
secondary indexes are used for the reasons explained in the previous section.

In queries 4 and 5 there are index lookups on \texttt{lineitem} using a primary
clustered index, and the cost of a lookup is 15us for the AppDirect and
30us Default mode. The lookup times are close, because the queries have a
smaller working set that mainly fits in the buffer cache. In general, for
smaller working sets secondary and primary index lookups have comparable lookup
times since they fit in the buffer cache and the OS page cache is not useful.
We also observe this behaviour in query 7.

In query 13, both modes take the same amount of time, because the working set is
larger than the buffer cache but smaller than the page cache. Thus, a memory
copy from DRAM to DRAM has similar latency to a memory copy from PMEM to DRAM.
Queries 19 and 20 involve a join with \texttt{lineitem} using a secondary index
lookup. Every lookup returns 30 and 7 tuples, respectively. PMEM is faster,
because of large block random accesses, even when a
small number of tuples is returned within a lookup.

Finally, query 21 has 3 joins with
\texttt{lineitem}. The first join has a barrier and there is little time
difference between the modes, because the index lookup happens in the clustered
primary key (i.e. \texttt{order\_key}) and therefore there are a lot of page
cache hits and prefetching is effective. Then, there is a second hash join with
\texttt{nation} that breaks the ordering of \texttt{order\_key}. That makes the
second join very slow in Default mode because the index lookups on the primary
key are random and the working set is very large (i.e. the entire
\texttt{lineitem}). Since the second join has no barrier, the third join happens
simultaneously with all the pages needed available in main memory. As a result,
for this join both modes take the same amount of time.

\subsubsection{Summary} We observe that operations that have large block
accesses ($\geq$ 16KB) index lookups with random accesses have a large
performance advantage in AppDirect mode compared to the Default mode. These
accesses cannot be effectively prefetched and the PMEM
bandwidth is very similar to a sequential access. Additionally, whenever
possible, we should use clustered indexes, since the additional lookup in a
non-clustered index has a performance penalty for PMEM due to the lack of an OS
page cache. For the Memory mode, we make similar conclusions to PostgreSQL,
i.e. that we should use it for queries with  working sets larger than the DRAM
available to take advantage of the larger page cache.

\subsection{TPC-C}
\subsubsection{Setup}
We use a large buffer pool (48GB unless otherwise mentioned) with similar
reasoning as in
Section~\ref{subsubsec:tpcc-postgres}. We disable binary logs and we leave the log
buffer size and the log file size at their default values, which is very small.
This causes checkpointing every second.

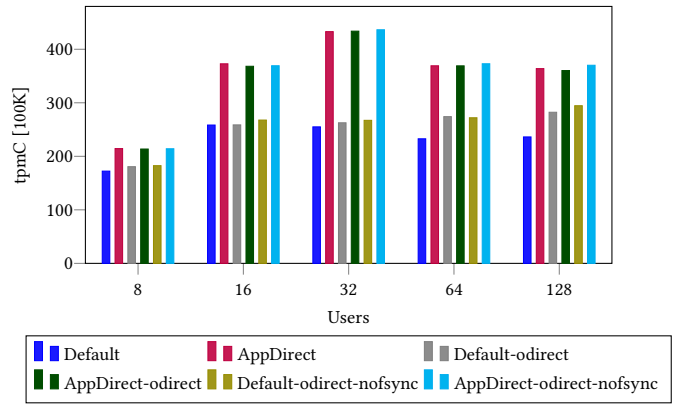
\begin{figure}[t]
  \begin{adjustwidth*}{-1.4em}{}
  \centering
  \input{plots/mysql-tpcc-flushing}
  \end{adjustwidth*}
  \caption{tpmC for TPC-C on MySQL with different flushing
  methods}\label{fig:tpcc-flushing-mysql}
  \vspace{-1em}
\end{figure}

\subsubsection{Flushing methods}
We first experiment with different flushing methods of InnoDB. More
specifically, we enable the \texttt{O\_DIRECT} flag, which uses direct I/O for
data files and buffered I/O for logs. We present the
results in Figure~\ref{fig:tpcc-flushing-mysql}. In general, the AppDirect mode is
faster than the Default mode, both with and without direct I/O, due to the
smaller latency that PMEM offers. However, direct I/O does not offer a
significant performance advantage in the AppDirect mode, because it does not
provide any additional benefit on top of DAX. Furthermore, using the AppDirect
mode with the \texttt{no-dax} option and the \texttt{O\_DIRECT} flag, provides
the same performance as the default AppDirect mode (with \texttt{dax} enabled).
This is expected, since both configurations skip the OS page cache.
For the Default mode, when we have
a small number of clients this happens because MySQL manages its own buffer pool
and does not depend upon the OS page cache for prefetching and caching. As the
number of clients increases, the Default mode with direct I/O performs much
better, since we have restricted the execution to one socket. Thus, the database
is not doing unnecessary memory copy from/to the OS page cache, which in turn
reduces contention and leaves more resources available. Based on these, we use
the \texttt{O\_DIRECT} flag for the rest of our experiments.

We also investigate how disabling \texttt{fsync} affects performance. This
function is used to ensure that writes are flushed to disk and not to the device
cache. Because these calls may degrade performance, MySQL provides a flushing
method called \texttt{O\_DIRECT\_NO\_FSYNC} that tries to take away these calls.
As we can see in Figure~\ref{fig:tpcc-flushing-mysql}, when we use this flag,
the performance increase is negligible, especially in AppDirect mode. This is a
result of mainly two reasons. Firstly, \texttt{fsync} calls are not totally
eliminated and they are still used for synchronizing file system metadata
changes, for example during appends~\cite{mysqlappends}. Secondly, writing dirty
pages to the tablespace is done in batches that require a single \texttt{fsync}
call that happens in the background. Therefore, this does not affect
transactions directly~\cite{fsyncbatch}. Contrary to the Default mode, when we disable
\texttt{fsync} we see a large improvement in the AppDirect mode. This happens
because overwrites in AppDirect mode with DAX use non-temporal stores. Thus, no
cache lines are flushed and only a \texttt{sfence} instruction is issued to
make sure the writes are complete.

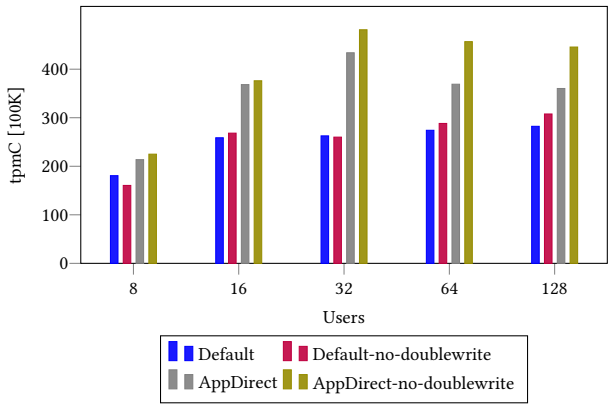
\begin{figure}[t]
  \begin{adjustwidth*}{-1.4em}{}
  \centering
  \input{plots/mysql-tpcc-concurrent-doublewrite}
  \end{adjustwidth*}
  \caption{tpmC for TPC-C on MySQL for different number of clients with and
  without the double-write buffer}\label{fig:tpcc-concurrent-doublewrite-mysql}
  \vspace{-1em}
\end{figure}

\subsubsection{Concurrent and double-write buffer}
We now experiment with the number of concurrent clients and the use of the
double-write buffer. By default, the double-write buffer is enabled both in the
Default and the AppDirect mode. We present the results in
Figure~\ref{fig:tpcc-concurrent-doublewrite-mysql}. For the Default mode, as the
number of concurrent clients increases, the tpmC first increases and then
reaches a plateau. Since the CPU utilization is low, the bottleneck is the SSD.
The latency keeps increasing due to increasing I/O and CPU scheduling delays. In
AppDirect mode, the tpmC increases and reaches a peak before dropping off again.
Given that we don't observe this behaviour in the Default mode, the drop is not
because of MySQL, but due to PMEM. This behaviour is consistent with other
studies~\cite{daase2021maximizing}, which notice a decrease in read and write
performance as the number of concurrent accesses increases. When the number of
read/writes operations increases, the read and write combining cache does not
perform so well, as its size is limited. Another reason for the drop in tpmC is
the increasing contention at the iMC level. When clients increase, there are
more L3 cache misses, which in turn lead to more DRAM accesses and I/O between
DRAM and PMEM.

The double-write buffer is a storage area where dirty pages are flushed to,
before writing to their respective positions in the data
files~\cite{doublewrite}. This buffer does not incur any additional cost,
because InnoDB writes the data in a large sequential chunk with a single
\texttt{fsync} call at the end. We disable the double-write buffer and present
the results in Figure~\ref{fig:tpcc-concurrent-doublewrite-mysql} for the two
modes. We see in the figure that the tpmC for the Default mode stays stable,
since double writes are a part of the background flushing. Therefore, the
storage can follow along and transactions are not affected. Conversely,
the tpmC increases significantly for the AppDirect mode when double-writes are
disabled, confirming once more that concurrent writes are a pain point for PMEM.

\begin{figure}[t]
  \begin{adjustwidth*}{}{}
  \centering
  \input{plots/mysql-tpcc-write-threads}
  \end{adjustwidth*}
  \caption{tpmC for TPC-C on MySQL for different number of write
  threads}\label{fig:tpcc-mysql-write-threads}
  \vspace{-2em}
\end{figure}
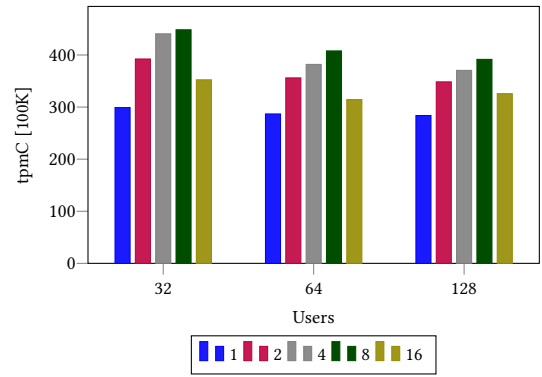

\subsubsection{I/O threads}
By default, MySQL uses async I/O to perform reads and writes. This removes the
control over how many concurrent requests are pending, which might be
detrimental to PMEM's performance. By using sync I/O, we can control the number
of background write threads. Read threads are mainly used for prefetching and
with a large buffer pool they don't have a significant role for the TPC-C
workload. We vary the write threads for different number of concurrent users and
we present the results in Figure~\ref{fig:tpcc-mysql-write-threads}. The throughput
reaches a maximum for 8 threads for different numbers of
concurrent users and decreases after that, confirming once again that PMEM is
not very effective at performing many concurrent accesses.

\begin{figure}[b]
  \begin{adjustwidth*}{}{}
  \centering
  \input{plots/mysql-tpcc-block-size}
  \end{adjustwidth*}
  \caption{tpmC for TPC-C on MySQL for different block sizes on
  AppDirect mode}\label{fig:tpcc-mysql-block-size}
  \vspace{-2em}
\end{figure}
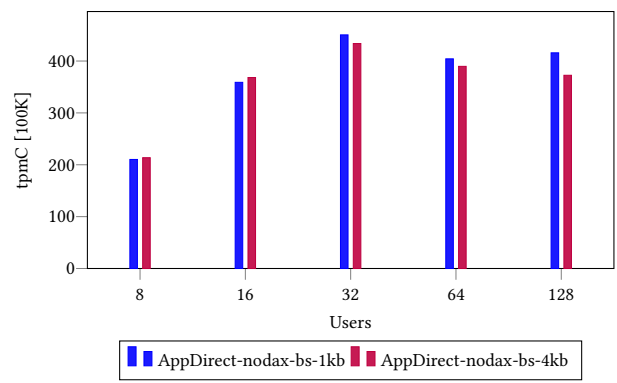

\subsubsection{ext4 block size}
We also evaluate how the ext4 block size affects PMEM's performance. We present
the results in Figure~\ref{fig:tpcc-mysql-block-size}. We set the ext4 block
size to 1KB and 4KB in AppDirect mode with \texttt{nodax} and \texttt{O\_DIRECT} enabled,
because direct access is only supported when the block size is equal to the
system page size. As we observe in the figure, for a small number of clients 1KB
block size performs worse than 4KB, because the amount of I/O is increased
and I/O is more in the critical path for a small number of clients. Contrary to
that, tpmC is larger for 1KB block size
for larger number of clients, because access size must be lower for higher
thread counts~\cite{daase2021maximizing}.

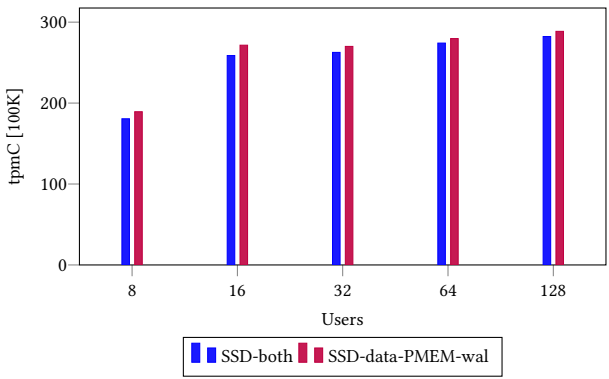
\begin{figure}[b]
  \begin{adjustwidth*}{}{}
  \centering
  \input{plots/mysql-tpcc-pmem-log}
  \end{adjustwidth*}
  \caption{tpmC for TPC-C on MySQL with placing the log directory in
  PMEM}\label{fig:tpcc-mysql-pmem-log}
\end{figure}

\subsubsection{Log placement}
To confirm that placing the logs in PMEM does not offer a significant performance
advantage irrespective of the DBMS, we place again the log directory in PMEM
instead of the SSD and we present the results in
Figure~\ref{fig:tpcc-mysql-pmem-log}. We can again confirm that redo logs are
not the bottleneck and buffered I/O hides the latency for the Default mode.


\subsubsection{Buffer pool size and app direct modes}
We finally experiment with the buffer pool size and different modes in AppDirect
mode (\texttt{no-dax}, \texttt{sector}). For brevity, we omit the figures and
we present only the high-level conclusions. With a smaller buffer pool (16GB),
we have more reads per transaction, while writes remain constant because of the
aggressive flushing mode. Thus, the AppDirect mode
remains unaffected since its read latency (10us) is negligible compared to the
overall query latency (10ms). On the other hand, for the Default mode when we
have a small number of clients ($\leq 16$) and a small buffer pool (16GB) the
tpmC drops around 20\%. This happens due to memory copies between the page cache
and the buffer pool. For large number of clients, having a larger
buffer pool (48GB) still increases tpmC but not as significantly, since time is
split more evenly between I/O and CPU. Regarding different configurations
in the AppDirect mode, we experiment with the \texttt{no-dax} and the
\texttt{sector} mode. Both perform worse than the \texttt{dax} mode, because we
have more levels of indirection. For the \texttt{no-dax} mode, reads and writes
still go through the block layer and this requires alignment verification. For
the \texttt{sector} mode, besides the block layer, we also have to go through
the block translation table.

\subsubsection{Summary}
Similar to Postgres, log placement in PMEM gives a negligible performance
advantage. Since concurrent writes are a bottleneck for PMEM, we should
finetune the respective parameters carefully. Specifically, increasing the
number of write threads above a threshold decreases performance since the
hardware is not as effective in write combining. We can further increase
performance by disabling the double-write buffer, which reduces concurrent
writes. Performance also drops when there is a lot of interference between
reads and writes, e.g. when the buffer pool has a small size. Finally, we see
again that in AppDirect mode \texttt{no-dax} performs worse than \texttt{fsdax}
because there are more levels of indirection.

\section{SQLServer}
\subsection{TPC-H}
\subsubsection{Setup}
We use SQLServer version 2019-15.0.4178. We set the buffer cache to 16GB to put
I/O in the critical path and have a consistent configuration across DBMSs.
We add non-clustered indexes to foreign keys as SQLServer does not do this
automatically. For the AppDirect mode we also place the
\texttt{tempdb} that stores intermediate results in PMEM. However, SQLServer
does not allow placing logs in a \texttt{fsdax}
filesystem~\cite{sqlserverpersistent}. Thus, in AppDirect mode we leave them in
the SSD. For OLAP workloads this does not affect performance.

\begin{figure}[t]
  \begin{adjustwidth*}{-1.5em}{}
  \centering
  \input{plots/sqlserver-tpch-runtime}
  \end{adjustwidth*}
  \caption{Running time (log-scale) for TPC-H SF-100 on SQLServer for different system
  configurations}\label{fig:tpch-time-sqlserver}
  \vspace{-1em}
\end{figure}
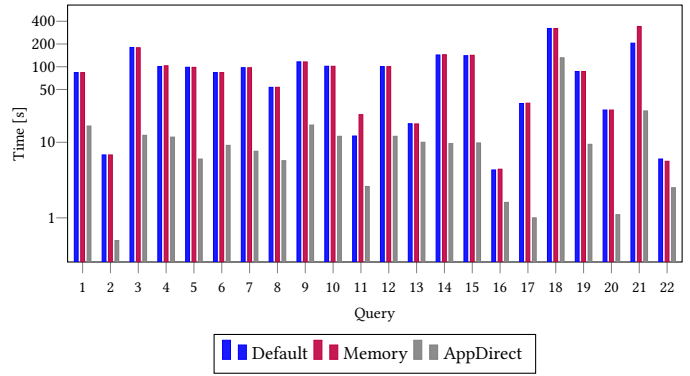
\subsubsection{General observations}
We present the running times for the 22 TPC-H queries for SQLServer for the
Default, Memory and AppDirect modes in Figure~\ref{fig:tpch-time-sqlserver}.
SQLServer uses in general all the available logical cores in socket 0 (48 in
total) and that is why the running time is much lower than the other two
databases. It is also very efficient at optimizing the queries and pipelining I/O
and processing. Therefore it reads only once the necessary tables for each query
and does not incur redundant I/O. Additionally, SQLServer prefers to write through
to physical hardware bypassing the OS page cache~\cite{sqlserverpagecache}.

The Default and the Memory mode have almost identical running times
for all but two queries. The average CPU utilization is very low for these two
modes, indicating that most of the time is spent transferring data from disk to
main memory. Prefetching is done by the iMC and processing is
interleaved with I/O very efficiently, making the working
set of almost all the queries to fit in the L4 cache, making essentially the two
modes running under the same hardware. This comes in contrast with the findings
of Wu et al.~\cite{wu2020lessons}, where the Memory mode was slower than the
Default mode. However, the authors of this paper perform warm runs, whereas in
our case we clear all the caches before performing measurements. That indicates
that the Memory mode may be advantageous when I/O is involved and PMEM is
effective at hiding the latency difference with DRAM in these cases. We are also
not aware of what other optimizations the authors of this paper might have used
leading to the observed performance difference. The two queries where Memory mode is
slower is queries 11 and 21, where
there are many conflict misses in the L4 DRAM cache and a significant data
transfer between DRAM and PMEM, compared to the Default mode. In addition to
that, as mentioned, SQLServer doesn't utilize the OS page cache and therefore does
not take advantage of the larger volatile memory available in the Memory mode.

Compared to the Default mode, the AppDirect mode is much faster again except for
one query. The difference is due to the lower latency that PMEM offers compared
to SSDs. Furthermore, since SQLServer uses many threads, it takes advantage of
the full PMEM bandwidth and it can reach read bandwidth up to 8 GB/s for some
queries. The difference is more obvious for queries involving the
\texttt{lineitem} table, since it is the larger table in the benchmark (around
100GB of storage together with indexes). We also notice that due to the very
high bandwidth SQL server had zero read-ahead reads, i.e. it doesn't place any
pages into the cache for the query.

\subsubsection{Query-by-query}
Since SQLServer is not open-source as the other two databases, we only have
estimates of the I/O costs and not the exact runtimes as in PostgreSQL and
MySQL. Thus, we cannot provide details about the application memory latency. We
choose again a subset of the queries with interesting insights and we compare
the Default with the AppDirect mode.

Query 5 has a clustered index seek on \texttt{lineitem} on \texttt{orderkey}.
\texttt{Lineitem} and \texttt{orders} are the two largest tables of TPC-H and
therefore after this seek the working set is reduced substantially. PMEM has
6$\times$ more read bandwidth than the SSD for this particular query and it
processes tuples as they come, without the need to put additional pages to the
buffer cache. The joins of the query are either nested
loop joins or hash joins. Since I/O is out of the critical path, there are more
resources to efficiently execute the joins resulting in an overall 16$\times$
faster runtime for the AppDirect mode compared to the Default mode.

Query 8 has a nested loop join with \texttt{lineitem} resulting in a lot of data
movement from/to the buffer cache in AppDirect mode. In contrast, this plan
puts the Default mode in a disadvantage as the DBMS
spends most of the time doing read-ahead reads to the buffer cache.
Compared to the other queries the CPU utilization of the AppDirect mode is low
(around 60\%, where for most of the other it is around 80\%).
Finally, query 18 is only 3$\times$ faster in AppDirect than in the Default mode. In
the query plan, we can see that there are 3 nested loop joins and 2 stream
aggregations on \texttt{lineitem} that expect sorted data. Therefore the query
is more CPU than I/O intensive, and that is the reason of the smaller
proportional time difference compared to the rest of the benchmark.

\subsubsection{Summary}
We see that the Memory mode behaves very similarly to the Default mode, except
for queries that cause conflict misses in the DRAM L4 cache. Since SQLServer
avoids using the OS page cache, it cannot take advantage of the larger
capacity of PMEM. On the other hand, the DBMS uses many threads for I/O and
processing and utilizes the full read bandwidth of PMEM, making the AppDirect mode
much faster than the other two modes.

\subsection{TPC-C}
\subsubsection{Setup}
We set the buffer cache to 48GB for similar reasons as in
Section~\ref{subsubsec:tpcc-postgres}. We do automatic checkpoints every minute
(the lowest possible value allowed by SQLServer). To use PMEM, SQLServer
requires \texttt{fsdax} mode to store the data and \texttt{sector} mode to store
the logs~\cite{sqlserverpersistent}. We therefore have to use both memory
sockets, as it is not possible to
create a mixed namespace in one socket that has the capacity for 1000
warehouses. We have observed that using PMEM in interleaved mode increases
performance for PMEM. Thus, we use only this mode in this section. Finally,
We also place the \texttt{temp} database and logs, which store
intermediate results, to the different storage mediums.
\subsubsection{Different modes and concurrent users}
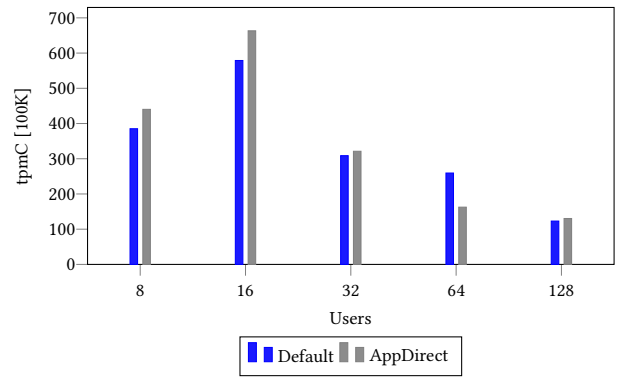
\begin{figure}[t]
  \begin{adjustwidth*}{-1.5em}{}
  \centering
  \input{plots/sqlserver-tpcc-users}
  \end{adjustwidth*}
  \caption{tpmC for TPC-C on SQLServer for different number of concurrent users
  }\label{fig:tpcc-modes-sqlserver}
  \vspace{-1em}
\end{figure}

We first experiment with the number of concurrent clients and we present the
results in Figure~\ref{fig:tpcc-modes-sqlserver}. In general, SQLServer uses
all the available cores in socket 0. As we can observe, for small
number of concurrent users, the higher bandwidth that PMEM offers gives a small
performance advantage compared to the Default mode, due to the lower latency it
provides compared to SSDs. However, as the number of concurrent users increases,
the Default mode is very competitive and outperforms the AppDirect mode for 64
concurrent users. For such a high number of users, there is a lot of interference
between I/O and processing in the system. Additionally, there is a large number
of concurrent reads and writes, which causes a significant performance drop for
PMEM~\cite{daase2021maximizing}. This performance reduction is so significant
that SSD outperforms PMEM for 64 users. Finally, we observe that CPU utilization is
very low for the AppDirect mode, indicating that I/O is in the critical path,
despite the much lower latency that PMEM offers.
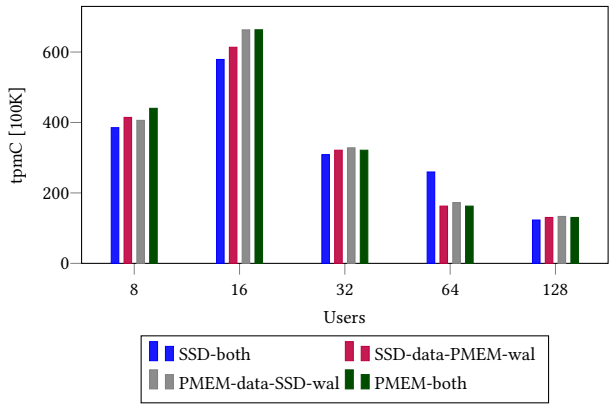
\begin{figure}[t]
  \begin{adjustwidth*}{-1.5em}{}
  \centering
  \input{plots/sqlserver-tpcc-stores}
  \end{adjustwidth*}
  \caption{tpmC for TPC-C on SQLServer using PMEM as WAL or data
  store}\label{fig:tpcc-stores-postgres}
  \vspace{-1em}
\end{figure}
\subsubsection{Log placement} We experiment with log placement as we did for the
other databases. Generally, placing the log in PMEM offers only
a small performance advantage, compared to placing the data in PMEM. This shows
that I/O is more on the critical path than the redo logs. As the number of
clients increases, we can see that log placement does not make a significant
difference. Lastly, placing the data on PMEM and the log
on SSD has a higher throughput than placing both on PMEM, but this is
due to the remote accesses to socket 1 for the log (that is necessary due to
the configuration enforced by SQLServer).

\subsubsection{Summary}
Since SQLserver utilizes all the available cores in the socket, the performance
drop when there are many concurrent reads and writes in PMEM is very obvious
compared to the other databases. We again confirm that the log placement does
not play a significant role in the overall performance, showing that PMEM is
more attractive for its lower latency rather than its persistence.

\section{Discussion}
We summarize the main insights from all the previous sections
and provide best practices for the different configurations that a
DBMS should follow to maximize the PMEM utilization.
\begin{enumerate}[label=(\roman*), noitemsep,topsep=0pt, leftmargin=1.2em]
    \item \textbf{Memory mode does not offer a significant performance advantage}:
         As we have observed for all databases and benchmarks, for the vast
         majority of cases, Memory mode with its larger volatile memory capacity
         does not offer any performance advantage. On the other hand, there are
         a lot of cases, where performance is slightly worse. This happens due
         to the conflict misses in the DRAM L4 cache, which also causes
         additional memory traffic between PMEM and DRAM.
    \item \textbf{Memory mode is useful for queries with sequential
        accesses and low number of reads/writes to memory}: In the case of
        queries with sequential accesses, the OS can take advantage of the
        larger filesystem cache offered by PMEM and together with prefetching
        these queries can have a significant speedup compared to
        configurations that do not use PMEM at all. If there is significant
        memory traffic between the different storage layers (e.g. when the
        workload performs many writes or the working set is larger than the OS
        page cache) the Memory mode can experience a small performance overhead.
    \item \textbf{Performance gains when using PMEM vs. SSDs can be due to
        application limitations rather than differences in hardware}: Although PMEM has
        superior performance compared to SSDs for sequential accesses with many
        threads, this is not the same for random accesses. If applications adopt
        asynchronous I/O for random accesses (e.g. index lookups) for workloads
        in which CPU requests overshadow I/O, the Default and the AppDirect mode will
        have a very small performance difference.
    \item \textbf{The lower latency of PMEM in AppDirect mode does not
        translate to an advantage equal to the hardware characteristics}:
        From our microbenchmarks, PMEM has 7-8$\times$ lower latency from an
        SSD and in some cases more than 10$\times$ higher read and write
        bandwidth. This difference does not translate to the expected runtime
        decrease as the I/O path is not fully optimized. The AppDirect mode with
        \texttt{fsdax}, which is the one recommended by Intel, does not have the
        advantage of an OS page cache. This feature is largely beneficial for
        sequential queries with high selectivity where prefetching is very
        efficient. It is also useful in the case of secondary indexes, where
        data from tables is stored in the page cache.
    \item \textbf{In systems where resources are limited, PMEM in AppDirect mode is not as
        useful}: In general PMEM in AppDirect mode involves the CPU as no DMA is
        available. When there is a lot of resource contention, the
        hardware advantage of PMEM is almost negligible. We noticed this
        behaviour in all three databases (in MySQL for TPC-H, when
        we restricted PostgreSQL to one core and in SQLServer for TPC-C).
    \item \textbf{PMEM requires fine-tuning for write-heavy workloads}: As
        noticed by previous
        studies~\cite{daase2021maximizing,izraelevitz2019basic} heavy-write
        workloads and read/write interference decrease the
        performance of PMEM dramatically. We noticed the same behaviour when
        running TPC-C. Therefore, we should carefully
        tune the number of write threads to avoid
        interference. Especially in the context of databases,
        several configurations that increase additional writes should be
        turned off to increase performance (e.g. the double-write buffer)
    \item \textbf{Optimizations made for SSDs/HDDs need to be re-designed when
        using PMEM}: Many traditional optimizations try to
        avoid I/O as much as possible due to the latency gap between main memory
        and storage (e.g. log compression). However, this gap is not as large
        with PMEM and these optimizations may not offer any performance advantage. In
        general, as the CPU is involved in I/O in AppDirect mode, it is preferable to
        avoid devoting CPU resources to optimize I/O.
    \item \textbf{Log placement in PMEM does not increase performance
        significantly}: As we have observed across databases, log placement in
        PMEM does not increase the transaction rate for TPC-C significantly.
        Placing data in PMEM increases throughout for TPC-C due to the lower
        latency and higher bandwidth of PMEM compared to SSD/HDD.
\end{enumerate}
\section{Conclusion}
In this paper, we benchmarked PMEM under three different relational engines (PostgreSQL,
MySQL and SQLServer) and two popular benchmarks (TPC-H and TPC-C). Our study
sheds light on how to efficiently integrate PMEM into the memory hierarchy and
how to tune DBMSs to get the best performance out of each configuration. In
Memory mode, increasing
volatile memory capacity using PMEM does not offer any
performance advantage. In AppDirect mode, PMEM
offers a lower latency than SSDs, which can increase performance significantly
when I/O is in the critical path. However, when there is a lot of resource
contention and because the I/O path is not fully optimized in the case of PMEM,
SSDs still remain a competitive solution for a number of workloads.

\bibliographystyle{ACM-Reference-Format}
\bibliography{sample}

\end{document}

%% file: figures/architecture.tex
\begin{tikzpicture}[node distance=1.5cm, auto]
    \node[cascaded] (cores1) {\small{Cores}};
    \node[cascaded, right of=cores1] (cores2) {\small{Cores}};
    \node[rectangle, thick, draw=black, below of=cores1, xshift=0.75cm, node
    distance=0.75cm, minimum width=3.5cm] (llc) {\small{LLC}};
    \node[rectangle, thick, draw=black, below of=llc, xshift=-1.5cm, node
    distance=0.5cm] (imc) {\small{iMC}};
    \node[rectangle, thick, draw=black, below of=llc, xshift=1.5cm, node
    distance=0.5cm] (imc2) {\small{iMC}};
    \node [draw, thick, inner xsep=0.5em, inner ysep=1.3em, fit=
        (cores1)(cores2)(llc)(imc)(imc2)] (socket) {};
    \node[fill=white] at (socket.north) {\footnotesize{Socket \#0}};
    \node [rectangle, below of=imc, node distance=1cm, scale=0.01] (aux) {};
    \node [rectangle, left of=aux, node distance=1cm, scale=0.01] (aux2) {};
    \node [rectangle, right of=aux, node distance=1cm, scale=0.01] (aux3) {};
    \node[rectangle, thick, draw=black, right of=socket, node
    distance=3.5cm] (ssd) {\small{SSD}};
    \node[rectangle, thick, draw=black, below of=aux, node
    distance=1cm] (dram1) {\footnotesize{DRAM}};
    \node[rectangle, thick, draw=black, left of=dram1, node
    distance=1cm] (pmem) {\footnotesize{DRAM}};
    \node[rectangle, thick, draw=black, right of=dram1, node
    distance=1cm] (dram2) {\footnotesize{PMEM}};
    \draw[-] (imc) -- (aux)  node [midway, left, yshift=-0.3cm] {
        \footnotesize{Memory Bus}};
    \node [rectangle, below of=imc2, node distance=1cm, scale=0.01] (aux4) {};
    \node [rectangle, left of=aux4, node distance=1cm, scale=0.01] (aux5) {};
    \node [rectangle, right of=aux4, node distance=1cm, scale=0.01] (aux6) {};
    \node[rectangle, thick, draw=black, below of=aux4, node
    distance=1cm] (dram3) {\footnotesize{DRAM}};
    \node[rectangle, thick, draw=black, right of=dram3, node
    distance=1cm] (pmem2) {\footnotesize{PMEM}};
    \node[rectangle, thick, draw=black, left of=dram3, node
    distance=1cm] (dram4) {\footnotesize{DRAM}};
    \draw[-] (imc2) -- (aux4)  node [midway, right, yshift=-0.3cm] {
        \footnotesize{Memory Bus}};
    \draw[-] (aux) -- (dram1);
    \draw[-] (aux) -- (aux3);
    \draw[-] (aux3) -- (dram2);
    \draw[-] (aux) -- (aux2);
    \draw[-] (aux2) -- (pmem);
    \draw[-] (aux4) -- (dram3);
    \draw[-] (aux4) -- (aux5);
    \draw[-] (aux5) -- (dram4);
    \draw[-] (aux4) -- (aux6);
    \draw[-] (aux6) -- (pmem2);
    \draw[-] (socket) -- (ssd) node [midway, below] {
        \footnotesize{PCIe Bus}};
\end{tikzpicture}

%% file: figures/memorymode.tex
\begin{tikzpicture}[auto]
    \node[isosceles triangle, thick, draw=black,
    rotate=90] (cpu) {
    \rotatebox{-90}{\scriptsize{CPU}}};
    \node[rectangle, thick, draw=black, below of=cpu,
    node distance=0.4cm, minimum width=4.1em] (l1) {\scriptsize{L1 cache}};
    \node[rectangle, thick, draw=black, below of=l1,
    node distance=0.4cm, minimum width=4.1em] (l2) {\scriptsize{L2 cache}};
    \node[rectangle, thick, draw=black, below of=l2,
    node distance=0.4cm, minimum width=4.1em] (l3) {\scriptsize{L3 cache}};
    \node[rectangle, thick, draw=black, below of=l3,
    node distance=0.6cm, minimum width=4.1em] (l4) {
    \scriptsize{\begin{tabular}{c}DRAM \\ (L4 cache)\end{tabular}}};
    \node[rectangle, thick, draw=black, below of=l4,
    node distance=0.75cm, minimum width=4.1em] (pmem) {
    \scriptsize{\begin{tabular}{c}PMEM \\ (Volatile)\end{tabular}}};
    \node[rectangle, thick, draw=black, below of=pmem,
    node distance=0.8cm, minimum width=4.1em] (ssd) {
    \scriptsize{\begin{tabular}{c}SSD \\ (Persistent)\end{tabular}}};
\end{tikzpicture}

%% file: figures/storagemode.tex
\begin{tikzpicture}[auto]
    \node[isosceles triangle, thick, draw=black,
    rotate=90] (cpu) {
    \rotatebox{-90}{\scriptsize{CPU}}};
    \node[rectangle, thick, draw=black, below of=cpu,
    node distance=0.4cm, minimum width=4.1em] (l1) {\scriptsize{L1 cache}};
    \node[rectangle, thick, draw=black, below of=l1,
    node distance=0.4cm, minimum width=4.1em] (l2) {\scriptsize{L2 cache}};
    \node[rectangle, thick, draw=black, below of=l2,
    node distance=0.4cm, minimum width=4.1em] (l3) {\scriptsize{L3 cache}};
    \node[rectangle, thick, draw=black, below of=l3,
    node distance=0.6cm, minimum width=4.1em] (l4) {
    \scriptsize{\begin{tabular}{c}DRAM \\ (Volatile)\end{tabular}}};
    \node[rectangle, thick, draw=black, below of=l4,
    node distance=0.75cm, minimum width=4.1em] (pmem) {
    \scriptsize{\begin{tabular}{c}PMEM \\ (Persistent)\end{tabular}}};
    \node[rectangle, thick, draw=black, below of=pmem,
    node distance=0.8cm, minimum width=4.1em] (ssd) {
    \scriptsize{\begin{tabular}{c}SSD \\ (Persistent)\end{tabular}}};
\end{tikzpicture}

%% file: figures/mixedmode.tex
\begin{tikzpicture}[auto]
    \node[isosceles triangle, thick, draw=black,
    rotate=90] (cpu) {
    \rotatebox{-90}{\scriptsize{CPU}}};
    \node[rectangle, thick, draw=black, below of=cpu,
    node distance=0.4cm, minimum width=4.1em] (l1) {\scriptsize{L1 cache}};
    \node[rectangle, thick, draw=black, below of=l1,
    node distance=0.4cm, minimum width=4.1em] (l2) {\scriptsize{L2 cache}};
    \node[rectangle, thick, draw=black, below of=l2,
    node distance=0.4cm, minimum width=4.1em] (l3) {\scriptsize{L3 cache}};
    \node[rectangle, thick, draw=black, below of=l3,
    node distance=0.6cm, minimum width=4.1em] (l4) {
    \scriptsize{\begin{tabular}{c}DRAM \\ (L4 cache)\end{tabular}}};
    \node[rectangle, thick, draw=black, below of=l4,
    node distance=0.75cm, xshift=-0.7cm, minimum width=4.1em] (pmem1) {
    \scriptsize{\begin{tabular}{c}PMEM \\ (Volatile)\end{tabular}}};
    \node[rectangle, thick, draw=black, below of=l4,
    node distance=0.75cm, xshift=0.7cm, minimum width=4.1em] (pmem2) {
    \scriptsize{\begin{tabular}{c}PMEM \\ (Persistent)\end{tabular}}};
    \node[rectangle, thick, draw=black, below of=l4,
    node distance=1.5cm, minimum width=4.1em] (ssd) {
    \scriptsize{\begin{tabular}{c}SSD \\ (Persistent)\end{tabular}}};
\end{tikzpicture}

%% file: plots/postgres-tpch-runtime.tex
\begin{tikzpicture}
\begin{axis}[
    ybar=1pt,
    ymode=log,
    log origin=infty,
    x=0.37cm,
    enlarge x limits={abs=0.2cm},
    ymin=0,
    ylabel shift=-1em,
    legend style={at={(0.5,-0.28)},
    anchor=north,legend columns=-1, font=\footnotesize},
    ylabel={Time [s]},
    xlabel={Query},
    symbolic x coords={1, 2, 3, 4, 5, 6, 7, 8, 9, 10, 11, 12, 13,
                       14, 15, 16, 17, 18, 19, 20, 21, 22},
    xtick=data,
    ytick={0, 10, 50, 100, 200, 300, 500, 1000},
    log ticks with fixed point,
    legend entries={Default, Memory, AppDirect},
    bar width=0.05cm,
    nodes near coords align={vertical},
    x tick label style={font=\scriptsize,text width=1cm,align=center},
    y tick label style={font=\scriptsize, xshift=0.1cm},
    tick align = outside,
    tick pos = left,
    ylabel near ticks,
    xlabel near ticks,
    ylabel style={font=\scriptsize},
    xlabel style={font=\scriptsize},
    every node near coord/.append style={font=\small},
    height=5cm,
    width=8cm
    ]
    \addplot[color=blue, fill=blue!90!white] coordinates {
        (1, 182) (2, 129) (3, 223) (4, 145) (5, 140) (6, 146)
        (7, 211) (8, 145) (9, 493) (10, 201) (11, 82) (12, 172)
        (13, 571) (14, 164) (15, 319) (16, 71) (17, 322) (18, 442)
        (19, 35) (20, 1760) (21, 208) (22, 9)} ;
    \addplot[color=purple, fill=purple!90!white] coordinates {
        (1, 182) (2, 138) (3, 216) (4, 147) (5, 142) (6, 147)
        (7, 216) (8, 142) (9, 285) (10, 204) (11, 88) (12, 175)
        (13, 667) (14, 165) (15, 241) (16, 73) (17, 326) (18, 436)
        (19, 37) (20, 1079) (21, 212) (22, 10)} ;
    \addplot[color=gray, fill=gray!90!white]  coordinates {
        (1, 106) (2, 73) (3, 47) (4, 20) (5, 21) (6, 18)
        (7, 49) (8, 22) (9, 63) (10, 48) (11, 27) (12, 34)
        (13, 440) (14, 27) (15, 61) (16, 72) (17, 242) (18, 290)
        (19, 6) (20, 156) (21, 48) (22, 6)} ;
\end{axis}
\end{tikzpicture}

%% file: plots/postgres-dram-read.tex
\begin{tikzpicture}
\begin{axis}[
    ybar=1pt,
    x=0.3cm,
    enlarge x limits={abs=0.2cm},
    ymin=0,
    legend style={at={(0.5,-0.28)},
    anchor=north,legend columns=-1, font=\footnotesize},
    ylabel={Bandwidth[MB/s]},
    xlabel={Query},
    symbolic x coords={1, 2, 3, 4, 5, 6, 7, 8, 9, 10, 11, 12, 13,
                       14, 15, 16, 17, 18, 19, 20, 21, 22},
    xtick=data,
    ytick={0, 250, 500, 750, 1000, 1250, 1500, 2500},
    legend entries={Default, Memory},
    bar width=0.05cm,
    nodes near coords align={vertical},
    x tick label style={font=\scriptsize,text width=1cm,align=center},
    y tick label style={font=\scriptsize, xshift=0.1cm},
    tick align = outside,
    tick pos = left,
    ylabel near ticks,
    xlabel near ticks,
    ylabel style={font=\scriptsize},
    xlabel style={font=\scriptsize},
    every node near coord/.append style={font=\small},
    height=5cm,
    width=8cm
    ]
    \addplot[color=blue, fill=blue!90!white] coordinates {
        (1, 316) (2, 404) (3, 718) (4, 596) (5, 711) (6, 372)
        (7, 650) (8, 785) (9, 832) (10, 593) (11, 318) (12, 526)
        (13, 834) (14, 354) (15, 619) (16, 240) (17, 231) (18, 1095)
        (19, 431) (20, 143) (21, 492) (22, 952)} ;
    \addplot[color=purple, fill=purple!90!white] coordinates {
        (1, 632) (2, 577) (3, 1104) (4, 1211) (5, 1311) (6, 779)
        (7, 1152) (8, 1462) (9, 2461) (10, 1160) (11, 497) (12, 1028)
        (13, 1061) (14, 758) (15, 1105) (16, 348) (17, 464) (18, 1290)
        (19, 980) (20, 427) (21, 966) (22, 870)} ;
\end{axis}
\end{tikzpicture}

%% file: plots/postgres-dram-write.tex
\begin{tikzpicture}
\begin{axis}[
    ybar=1pt,
    x=0.3cm,
    enlarge x limits={abs=0.2cm},
    ymin=0,
    legend style={at={(0.5,-0.28)},
    anchor=north,legend columns=-1, font=\footnotesize},
    ylabel={Bandwidth[MB/s]},
    xlabel={Query},
    symbolic x coords={1, 2, 3, 4, 5, 6, 7, 8, 9, 10, 11, 12, 13,
                       14, 15, 16, 17, 18, 19, 20, 21, 22},
    xtick=data,
    ytick={0, 250, 500, 750, 1000, 1250, 1500, 1750},
    legend entries={Default, Memory},
    bar width=0.05cm,
    nodes near coords align={vertical},
    x tick label style={font=\scriptsize,text width=1cm,align=center},
    y tick label style={font=\scriptsize, xshift=0.1cm},
    tick align = outside,
    tick pos = left,
    ylabel near ticks,
    xlabel near ticks,
    ylabel style={font=\scriptsize},
    xlabel style={font=\scriptsize},
    every node near coord/.append style={font=\small},
    height=5cm,
    width=8cm
    ]
    \addplot[color=blue, fill=blue!90!white] coordinates {
        (1, 562) (2, 427) (3, 851) (4, 1093) (5, 1050) (6, 718)
        (7, 894) (8, 1069) (9, 1260) (10, 939) (11, 416) (12, 919)
        (13, 604) (14, 704) (15, 778) (16, 271) (17, 373) (18, 822)
        (19, 885) (20, 240) (21, 770) (22, 974)} ;
    \addplot[color=purple, fill=purple!90!white] coordinates {
        (1, 807) (2, 589) (3, 1221) (4, 1741) (5, 1588) (6, 1065)
        (7, 1389) (8, 1595) (9, 1579) (10, 1455) (11, 579) (12, 1425)
        (13, 720) (14, 1008) (15, 1080) (16, 400) (17, 550) (18, 1031)
        (19, 1416) (20, 469) (21, 1227) (22, 1295)} ;
\end{axis}
\end{tikzpicture}

%% file: plots/postgres-read-storage.tex
\begin{tikzpicture}
\begin{axis}[
    ybar=1pt,
    x=0.37cm,
    enlarge x limits={abs=0.2cm},
    ymin=0,
    ylabel shift=-0.5em,
    legend style={at={(0.5,-0.28)},
    anchor=north,legend columns=-1, font=\footnotesize},
    ylabel={Data read [GB]},
    xlabel={Query},
    symbolic x coords={1, 2, 3, 4, 5, 6, 7, 8, 9, 10, 11, 12, 13,
                       14, 15, 16, 17, 18, 19, 20, 21, 22},
    xtick=data,
    ytick={0, 25, 50, 75, 100, 150, 250},
    legend entries={Default, Memory, AppDirect},
    bar width=0.05cm,
    nodes near coords align={vertical},
    x tick label style={font=\scriptsize,text width=1cm,align=center},
    y tick label style={font=\scriptsize, xshift=0.1cm},
    tick align = outside,
    tick pos = left,
    ylabel near ticks,
    xlabel near ticks,
    ylabel style={font=\scriptsize},
    xlabel style={font=\scriptsize},
    every node near coord/.append style={font=\small},
    height=5cm,
    width=8cm
    ]
    \addplot[color=blue, fill=blue!90!white] coordinates {
        (1, 85) (2, 18) (3, 113) (4, 84) (5, 71) (6, 85)
        (7, 110) (8, 60) (9, 262) (10, 93) (11, 14) (12, 108)
        (13, 21) (14, 88) (15, 171) (16, 7) (17, 90) (18, 101)
        (19, 13) (20, 173) (21, 103) (22, 3)} ;
    \addplot[color=purple, fill=purple!90!white] coordinates {
        (1, 85) (2, 18) (3, 108) (4, 83) (5, 71) (6, 85)
        (7, 109) (8, 57) (9, 128) (10, 93) (11, 14) (12, 107)
        (13, 21) (14, 88) (15, 85) (16, 7) (17, 89) (18, 89)
        (19, 13) (20, 108) (21, 104) (22, 3)} ;
    \addplot[color=gray, fill=gray!90!white]  coordinates {
        (1, 88) (2, 24) (3, 120) (4, 57) (5, 68) (6, 81)
        (7, 127) (8, 53) (9, 260) (10, 71) (11, 12) (12, 102)
        (13, 263) (14, 87) (15, 176) (16, 8) (17, 93) (18, 144)
        (19, 12) (20, 170) (21, 103) (22, 3)} ;
\end{axis}
\end{tikzpicture}

%% file: plots/postgres-tpcc-users.tex
\begin{tikzpicture}
\begin{axis}[
    ybar=2pt,
    x=1.4cm,
    enlarge x limits={abs=0.7cm},
    ymin=0,
    legend style={at={(0.5,-0.28)},
    anchor=north,legend columns=3, font=\footnotesize},
    ylabel={tpmC [100K]},
    xlabel={Users},
    symbolic x coords={8, 16, 32, 64, 128},
    xtick=data,
    ytick={0, 100, 200, 300, 400},
    legend entries={Default, Default-compress, AppDirect, AppDirect-nodax,
    AppDirect-compress},
    bar width=0.1cm,
    nodes near coords align={vertical},
    x tick label style={font=\footnotesize,text width=1cm,align=center},
    y tick label style={font=\footnotesize, xshift=0.1cm},
    tick align = outside,
    tick pos = left,
    ylabel near ticks,
    xlabel near ticks,
    ylabel style={font=\footnotesize},
    xlabel style={font=\footnotesize},
    every node near coord/.append style={font=\small},
    height=5cm,
    width=12cm
    ]
    \addplot[color=blue, fill=blue!90!white] coordinates {
        (8, 98.7) (16, 104) (32, 121) (64, 139) (128, 162.6)
    };
    \addplot[color=purple, fill=purple!90!white] coordinates {
        (8, 99.8) (16, 114) (32, 126) (64, 147) (128, 171.9)
    };
    \addplot[color=gray, fill=gray!90!white]  coordinates {
        (8, 319.3) (16, 428.9) (32, 438) (64, 431.4) (128, 429.6)
    };
    \addplot[color=green!30!black, fill=green!30!black]  coordinates {
        (8, 231.4) (16, 252) (32, 274.8) (64, 278.9) (128, 290.2)
    };
    \addplot[color=olive, fill=olive!90!white]  coordinates {
        (8, 105.7) (16, 221.2) (32, 380.5) (64, 388.9) (128, 372)
    };
\end{axis}
\end{tikzpicture}

%% file: plots/postgres-tpcc-dimms.tex
\begin{tikzpicture}
\begin{axis}[
    ybar=2pt,
    x=1.4cm,
    enlarge x limits={abs=0.7cm},
    ymin=0,
    legend style={at={(0.5,-0.28)},
    anchor=north,legend columns=3, font=\footnotesize},
    ylabel={tpmC [100K]},
    xlabel={Users},
    symbolic x coords={8, 16, 32, 64, 128},
    xtick=data,
    ytick={0, 200, 400, 600, 800, 1000, 1200},
    legend entries={AppDirect, AppDirect-interleaved,
    AppDirect-interleaved-increased-wal},
    bar width=0.1cm,
    nodes near coords align={vertical},
    x tick label style={font=\footnotesize,text width=1cm,align=center},
    y tick label style={font=\footnotesize, xshift=0.1cm},
    tick align = outside,
    tick pos = left,
    ylabel near ticks,
    xlabel near ticks,
    ylabel style={font=\footnotesize},
    xlabel style={font=\footnotesize},
    every node near coord/.append style={font=\small},
    height=5cm,
    width=12cm
    ]
    \addplot[color=blue, fill=blue!90!white] coordinates {
        (8, 319.3) (16, 428.9) (32, 438) (64, 431.4) (128, 429.6)
    };
    \addplot[color=purple, fill=purple!90!white] coordinates {
        (8, 346.8) (16, 524.2) (32, 524.6) (64, 533.9) (128, 477.4)
    };
    \addplot[color=gray, fill=gray!90!white]  coordinates {
        (8, 356.6) (16, 578) (32, 831.7) (64, 1029.1) (128, 1132)
    };
\end{axis}
\end{tikzpicture}

%% file: plots/postgres-tpcc-stores.tex
\begin{tikzpicture}
\begin{axis}[
    ybar=2pt,
    x=1.4cm,
    enlarge x limits={abs=0.7cm},
    ymin=0,
    legend style={at={(0.5,-0.28)},
    anchor=north,legend columns=2, font=\footnotesize},
    ylabel={tpmC [100K]},
    xlabel={Users},
    symbolic x coords={8, 16, 32, 64, 128},
    xtick=data,
    ytick={0, 200, 400, 600, 800, 1000, 1200},
    legend entries={SSD-both, SSD-data-PMEM-wal,
    PMEM-data-SSD-wal, PMEM-both},
    bar width=0.1cm,
    legend cell align={left},
    nodes near coords align={vertical},
    x tick label style={font=\footnotesize,text width=1cm,align=center},
    y tick label style={font=\footnotesize, xshift=0.1cm},
    tick align = outside,
    tick pos = left,
    ylabel near ticks,
    xlabel near ticks,
    ylabel style={font=\footnotesize},
    xlabel style={font=\footnotesize},
    every node near coord/.append style={font=\small},
    height=5cm,
    width=12cm
    ]
    \addplot[color=blue, fill=blue!90!white] coordinates {
        (8, 170.8) (16, 239.8) (32, 332.7) (64, 387) (128, 474.4)
    };
    \addplot[color=purple, fill=purple!90!white] coordinates {
        (8, 210.3) (16, 313.3) (32, 373.5) (64, 502.6) (128, 495.9)
    };
    \addplot[color=gray, fill=gray!90!white]  coordinates {
        (8, 324.4) (16, 520.1) (32, 541.6) (64, 598.7) (128, 645.7)
    };
    \addplot[color=green!30!black, fill=green!30!black]  coordinates {
        (8, 350.6) (16, 577.4) (32, 854.2) (64, 1001.7) (128, 1088.4)
    };
\end{axis}
\end{tikzpicture}

%% file: plots/mysql-tpch-runtime.tex
\begin{tikzpicture}
\begin{axis}[
    ybar=1pt,
    ymode=log,
    log origin=infty,
    x=0.37cm,
    enlarge x limits={abs=0.2cm},
    ymin=0,
    ylabel shift=-1em,
    legend style={at={(0.5,-0.28)},
    anchor=north,legend columns=-1, font=\footnotesize},
    ylabel={Time [s]},
    xlabel={Query},
    symbolic x coords={1, 2, 3, 4, 5, 6, 7, 8, 9, 10, 11, 12, 13,
                       14, 15, 16, 17, 18, 19, 20, 21, 22},
    xtick=data,
    ytick={0, 50, 100, 200, 300, 500, 1000, 1500, 2200, 3500},
    log ticks with fixed point,
    legend entries={Default, Memory, AppDirect},
    bar width=0.05cm,
    nodes near coords align={vertical},
    x tick label style={font=\scriptsize,text width=1cm,align=center},
    y tick label style={font=\scriptsize, xshift=0.1cm},
    tick align = outside,
    tick pos = left,
    ylabel near ticks,
    xlabel near ticks,
    ylabel style={font=\scriptsize},
    xlabel style={font=\scriptsize},
    every node near coord/.append style={font=\small},
    height=5cm,
    width=8cm
    ]
    \addplot[color=blue, fill=blue!90!white] coordinates {
        (1, 2273) (2, 36) (3, 2597) (4, 265) (5, 443) (6, 429)
        (7, 1203) (8, 769) (9, 5074) (10, 587) (11, 163) (12, 599)
        (13, 3457) (14, 484) (15, 1240) (16, 168) (17, 210) (18, 655)
        (19, 206) (20, 671) (21, 1950) (22, 50)} ;
    \addplot[color=purple, fill=purple!90!white] coordinates {
        (1, 2280) (2, 40) (3, 1494) (4, 273) (5, 462) (6, 442)
        (7, 805) (8, 804) (9, 2384) (10, 590) (11, 174) (12, 610)
        (13, 3796) (14, 502) (15, 1253) (16, 179) (17, 220) (18, 668)
        (19, 218) (20, 621) (21, 1571) (22, 53)} ;
    \addplot[color=gray, fill=gray!90!white]  coordinates {
        (1, 2241) (2, 22) (3, 939) (4, 183) (5, 340) (6, 392)
        (7, 364) (8, 688) (9, 1638) (10, 499) (11, 110) (12, 555)
        (13, 3442) (14, 426) (15, 1149) (16, 86) (17, 129) (18, 608)
        (19, 61) (20, 162) (21, 1520) (22, 32)} ;
\end{axis}
\end{tikzpicture}

%% file: plots/mysql-tpcc-flushing.tex
\begin{tikzpicture}
\begin{axis}[
    ybar=2pt,
    x=1.4cm,
    enlarge x limits={abs=0.7cm},
    ymin=0,
    legend style={at={(0.5,-0.28)},
    anchor=north,legend columns=3, font=\footnotesize},
    ylabel={tpmC [100K]},
    xlabel={Users},
    symbolic x coords={8, 16, 32, 64, 128},
    xtick=data,
    legend cell align={left},
    ytick={0, 100, 200, 300, 400},
    legend entries={Default, AppDirect, Default-odirect, AppDirect-odirect,
    Default-odirect-nofsync, AppDirect-odirect-nofsync},
    bar width=0.1cm,
    nodes near coords align={vertical},
    x tick label style={font=\footnotesize,text width=1cm,align=center},
    y tick label style={font=\footnotesize, xshift=0.1cm},
    tick align = outside,
    tick pos = left,
    ylabel near ticks,
    xlabel near ticks,
    ylabel style={font=\footnotesize},
    xlabel style={font=\footnotesize},
    every node near coord/.append style={font=\small},
    height=5cm,
    width=12cm
    ]
    \addplot[color=blue, fill=blue!90!white] coordinates {
        (8, 172.3) (16, 258.4) (32, 255) (64, 232.7) (128, 236.2)
    };
    \addplot[color=purple, fill=purple!90!white] coordinates {
        (8, 214.4) (16, 372.8) (32, 432.9) (64, 369.1) (128, 363.8)
    };
    \addplot[color=gray, fill=gray!90!white]  coordinates {
        (8, 180.6) (16, 258.7) (32, 262.6) (64, 274.1) (128, 282.3)
    };
    \addplot[color=green!30!black, fill=green!30!black]  coordinates {
        (8, 213.6) (16, 368.2) (32, 433.8) (64, 369) (128, 360.2)
    };
    \addplot[color=olive, fill=olive!90!white]  coordinates {
        (8, 182.5) (16, 267.7) (32, 267.2) (64, 272) (128, 294.5)
    };
    \addplot[color=cyan, fill=cyan!90!white]  coordinates {
        (8, 214.1) (16, 369.2) (32, 436.4) (64, 372.9) (128, 370.1)
    };
\end{axis}
\end{tikzpicture}

%% file: plots/mysql-tpcc-concurrent-doublewrite.tex
\begin{tikzpicture}
\begin{axis}[
    ybar=2pt,
    x=1.4cm,
    enlarge x limits={abs=0.7cm},
    ymin=0,
    legend style={at={(0.5,-0.28)},
    anchor=north,legend columns=2, font=\footnotesize},
    ylabel={tpmC [100K]},
    xlabel={Users},
    symbolic x coords={8, 16, 32, 64, 128},
    xtick=data,
    legend cell align={left},
    ytick={0, 100, 200, 300, 400},
    legend entries={Default, Default-no-doublewrite, AppDirect,
    AppDirect-no-doublewrite},
    bar width=0.1cm,
    nodes near coords align={vertical},
    x tick label style={font=\footnotesize,text width=1cm,align=center},
    y tick label style={font=\footnotesize, xshift=0.1cm},
    tick align = outside,
    tick pos = left,
    ylabel near ticks,
    xlabel near ticks,
    ylabel style={font=\footnotesize},
    xlabel style={font=\footnotesize},
    every node near coord/.append style={font=\small},
    height=5cm,
    width=12cm
    ]
    \addplot[color=blue, fill=blue!90!white]  coordinates {
        (8, 180.6) (16, 258.7) (32, 262.6) (64, 274.1) (128, 282.3)
    };
    \addplot[color=purple, fill=purple!90!white] coordinates {
        (8, 160.5) (16, 268.2) (32, 260) (64, 288.2) (128, 307.7)
    };
    \addplot[color=gray, fill=gray!90!white]  coordinates {
        (8, 213.6) (16, 368.2) (32, 433.8) (64, 369) (128, 360.2)
    };
    \addplot[color=olive, fill=olive!90!white]  coordinates {
        (8, 224.9) (16, 376.1) (32, 481.2) (64, 456.7) (128, 445.6)
    };
\end{axis}
\end{tikzpicture}

%% file: plots/mysql-tpcc-write-threads.tex
\begin{tikzpicture}
\begin{axis}[
    ybar=2pt,
    x=2cm,
    enlarge x limits={abs=1cm},
    ymin=0,
    legend style={at={(0.5,-0.28)},
    anchor=north,legend columns=-1, font=\footnotesize},
    ylabel={tpmC [100K]},
    xlabel={Users},
    symbolic x coords={32, 64, 128},
    xtick=data,
    legend cell align={left},
    ytick={0, 100, 200, 300, 400, 500},
    legend entries={1, 2, 4, 8, 16},
    bar width=0.2cm,
    nodes near coords align={vertical},
    x tick label style={font=\footnotesize,text width=1cm,align=center},
    y tick label style={font=\footnotesize, xshift=0.1cm},
    tick align = outside,
    tick pos = left,
    ylabel near ticks,
    xlabel near ticks,
    ylabel style={font=\footnotesize},
    xlabel style={font=\footnotesize},
    every node near coord/.append style={font=\small},
    height=5cm,
    width=12cm
    ]
    \addplot[color=blue, fill=blue!90!white]  coordinates {
        (32, 299) (64, 286.8) (128, 283.8)
    };
    \addplot[color=purple, fill=purple!90!white] coordinates {
        (32, 392.3) (64, 356) (128, 348.4)
    };
    \addplot[color=gray, fill=gray!90!white]  coordinates {
        (32, 440.6) (64, 381.9) (128, 370.4)
    };
    \addplot[color=green!30!black, fill=green!30!black]  coordinates {
        (32, 448.4) (64, 407.8) (128, 391.6)
    };
    \addplot[color=olive, fill=olive!90!white]  coordinates {
        (32, 352.3) (64, 314.3) (128, 325.7)
    };
\end{axis}
\end{tikzpicture}

%% file: plots/mysql-tpcc-block-size.tex
\begin{tikzpicture}
\begin{axis}[
    ybar=2pt,
    x=1.4cm,
    enlarge x limits={abs=0.7cm},
    ymin=0,
    legend style={at={(0.5,-0.28)},
    anchor=north,legend columns=2, font=\footnotesize},
    ylabel={tpmC [100K]},
    xlabel={Users},
    symbolic x coords={8, 16, 32, 64, 128},
    xtick=data,
    legend cell align={left},
    ytick={0, 100, 200, 300, 400},
    legend entries={AppDirect-nodax-bs-1kb, AppDirect-nodax-bs-4kb},
    bar width=0.1cm,
    nodes near coords align={vertical},
    x tick label style={font=\footnotesize,text width=1cm,align=center},
    y tick label style={font=\footnotesize, xshift=0.1cm},
    tick align = outside,
    tick pos = left,
    ylabel near ticks,
    xlabel near ticks,
    ylabel style={font=\footnotesize},
    xlabel style={font=\footnotesize},
    every node near coord/.append style={font=\small},
    height=5cm,
    width=12cm
    ]
    \addplot[color=blue, fill=blue!90!white]  coordinates {
        (8, 210.2) (16, 358.9) (32, 450.4) (64, 404.2) (128, 415.8)
    };
    \addplot[color=purple, fill=purple!90!white] coordinates {
        (8, 213.6) (16, 368.2) (32, 433.7) (64, 389.7) (128, 372.5)
    };
\end{axis}
\end{tikzpicture}

%% file: plots/mysql-tpcc-pmem-log.tex
\begin{tikzpicture}
\begin{axis}[
    ybar=2pt,
    x=1.4cm,
    enlarge x limits={abs=0.7cm},
    ymin=0,
    legend style={at={(0.5,-0.28)},
    anchor=north,legend columns=2, font=\footnotesize},
    ylabel={tpmC [100K]},
    xlabel={Users},
    symbolic x coords={8, 16, 32, 64, 128},
    xtick=data,
    legend cell align={left},
    ytick={0, 100, 200, 300, 400},
    legend entries={SSD-both, SSD-data-PMEM-wal},
    bar width=0.1cm,
    nodes near coords align={vertical},
    x tick label style={font=\footnotesize,text width=1cm,align=center},
    y tick label style={font=\footnotesize, xshift=0.1cm},
    tick align = outside,
    tick pos = left,
    ylabel near ticks,
    xlabel near ticks,
    ylabel style={font=\footnotesize},
    xlabel style={font=\footnotesize},
    every node near coord/.append style={font=\small},
    height=5cm,
    width=12cm
    ]
    \addplot[color=blue, fill=blue!90!white]  coordinates {
        (8, 180.6) (16, 258.7) (32, 262.6) (64, 274.1) (128, 282.2)
    };
    \addplot[color=purple, fill=purple!90!white] coordinates {
        (8, 189.3) (16, 271.5) (32, 270) (64, 279.7) (128, 288.6)
    };
\end{axis}
\end{tikzpicture}

%% file: plots/sqlserver-tpch-runtime.tex
\begin{tikzpicture}
\begin{axis}[
    ybar=1pt,
    x=0.37cm,
    ymode=log,
    log origin=infty,
    enlarge x limits={abs=0.2cm},
    ymin=0,
    ylabel shift=-0.5em,
    legend style={at={(0.5,-0.28)},
    anchor=north,legend columns=-1, font=\footnotesize},
    ylabel={Time [s]},
    xlabel={Query},
    symbolic x coords={1, 2, 3, 4, 5, 6, 7, 8, 9, 10, 11, 12, 13,
                       14, 15, 16, 17, 18, 19, 20, 21, 22},
    xtick=data,
    ytick={0, 1, 10, 50, 100, 200, 400},
    log ticks with fixed point,
    legend entries={Default, Memory, AppDirect},
    bar width=0.05cm,
    nodes near coords align={vertical},
    x tick label style={font=\scriptsize,text width=1cm,align=center},
    y tick label style={font=\scriptsize, xshift=0.1cm},
    tick align = outside,
    tick pos = left,
    ylabel near ticks,
    xlabel near ticks,
    ylabel style={font=\scriptsize},
    xlabel style={font=\scriptsize},
    every node near coord/.append style={font=\small},
    height=5cm,
    width=8cm
    ]
    \addplot[color=blue, fill=blue!90!white] coordinates {
        (1, 83.7) (2, 6.8) (3, 179.5) (4, 100.6) (5, 98.3) (6, 83.7)
        (7, 97.1) (8, 53.3) (9, 116) (10, 101.4) (11, 12.1) (12, 100.4)
        (13, 17.6) (14, 143) (15, 139.9) (16, 4.3) (17, 32.6) (18, 321)
        (19, 86.4) (20, 26.7) (21, 204.8) (22, 6)} ;
    \addplot[color=purple, fill=purple!90!white] coordinates {
        (1, 83.6) (2, 6.8) (3, 178.4) (4, 103) (5, 98.1) (6, 83.6)
        (7, 97) (8, 53.4) (9, 115.7) (10, 101.3) (11, 23.2) (12, 100.3)
        (13, 17.5) (14, 144.2) (15, 141.7) (16, 4.4) (17, 32.9) (18, 320.4)
        (19, 86.4) (20, 26.7) (21, 341.2) (22, 5.6)} ;
    \addplot[color=gray, fill=gray!90!white]  coordinates {
        (1, 16.4) (2, 0.5) (3, 12.36) (4, 11.7) (5, 6) (6, 9.1)
        (7, 7.6) (8, 5.7) (9, 16.9) (10, 12) (11, 2.6) (12, 12)
        (13, 10) (14, 9.6) (15, 9.8) (16, 1.6) (17, 1) (18, 131.3)
        (19, 9.4) (20, 1.1) (21, 26) (22, 2.5)} ;
\end{axis}
\end{tikzpicture}

%% file: plots/sqlserver-tpcc-users.tex
\begin{tikzpicture}
\begin{axis}[
    ybar=2pt,
    x=1.4cm,
    enlarge x limits={abs=0.7cm},
    ymin=0,
    legend style={at={(0.5,-0.28)},
    anchor=north,legend columns=3, font=\footnotesize},
    ylabel={tpmC [100K]},
    xlabel={Users},
    symbolic x coords={8, 16, 32, 64, 128},
    xtick=data,
    ytick={0, 100, 200, 300, 400, 500, 600, 700},
    legend entries={Default, AppDirect},
    bar width=0.1cm,
    nodes near coords align={vertical},
    x tick label style={font=\footnotesize,text width=1cm,align=center},
    y tick label style={font=\footnotesize, xshift=0.1cm},
    tick align = outside,
    tick pos = left,
    ylabel near ticks,
    xlabel near ticks,
    ylabel style={font=\footnotesize},
    xlabel style={font=\footnotesize},
    every node near coord/.append style={font=\small},
    height=5cm,
    width=12cm
    ]
    \addplot[color=blue, fill=blue!90!white] coordinates {
        (8, 385.2) (16, 578.9) (32, 308.7) (64, 259.4) (128, 123)
    };
    \addplot[color=gray, fill=gray!90!white]  coordinates {
        (8, 440.2) (16, 663.2) (32, 321.2) (64, 162.5) (128, 130.3)
    };
\end{axis}
\end{tikzpicture}

%% file: plots/sqlserver-tpcc-stores.tex
\begin{tikzpicture}
\begin{axis}[
    ybar=2pt,
    x=1.4cm,
    enlarge x limits={abs=0.7cm},
    ymin=0,
    legend style={at={(0.5,-0.28)},
    anchor=north,legend columns=2, font=\footnotesize},
    ylabel={tpmC [100K]},
    xlabel={Users},
    symbolic x coords={8, 16, 32, 64, 128},
    xtick=data,
    ytick={0, 200, 400, 600, 800, 1000, 1200},
    legend entries={SSD-both, SSD-data-PMEM-wal,
    PMEM-data-SSD-wal, PMEM-both},
    bar width=0.1cm,
    legend cell align={left},
    nodes near coords align={vertical},
    x tick label style={font=\footnotesize,text width=1cm,align=center},
    y tick label style={font=\footnotesize, xshift=0.1cm},
    tick align = outside,
    tick pos = left,
    ylabel near ticks,
    xlabel near ticks,
    ylabel style={font=\footnotesize},
    xlabel style={font=\footnotesize},
    every node near coord/.append style={font=\small},
    height=5cm,
    width=12cm
    ]
    \addplot[color=blue, fill=blue!90!white] coordinates {
        (8, 385.2) (16, 578.9) (32, 308.7) (64, 259.4) (128, 123)
    };
    \addplot[color=purple, fill=purple!90!white] coordinates {
        (8, 414.3) (16, 613.5) (32, 321.2) (64, 162.5) (128, 130.3)
    };
    \addplot[color=gray, fill=gray!90!white]  coordinates {
        (8, 406) (16, 662.9) (32, 328) (64, 172.6) (128, 133.2)
    };
    \addplot[color=green!30!black, fill=green!30!black]  coordinates {
        (8, 440.2) (16, 663.2) (32, 321.2) (64, 162.5) (128, 130.3)
    };
\end{axis}
\end{tikzpicture}